\documentclass[draftcls,12pt,onecolumn]{IEEEtran}

\usepackage{authblk}
\usepackage{amsmath}
\usepackage{amssymb}
\usepackage{graphicx}
\usepackage{array}
\usepackage{makecell}
\usepackage{tikz}
\usepackage{pgfplots}	

\usepackage{booktabs}
\usepackage{floatflt}
\usepackage{enumerate}
\usepackage{psfrag}	
\usepackage{array}
\usepackage{multirow,hhline}
\usepackage{exscale}
\usepackage{color}
\usepackage{colortbl}
\usepackage{epsfig,subfigure}
\usepackage{cite}
\usepackage{amsthm}
\usepackage{algorithm}
\usepackage{algpseudocode}
\usepackage[font=small]{caption}
\usepackage[latin1]{inputenc} 
\usepackage[T1]{fontenc} 
\usepackage{enumerate}

\newcommand{\norm}[1]{\left\lVert#1\right\rVert}

\newtheorem{definition}{Definition}
\newtheorem{lemma}{Lemma}
\newtheorem{remark}{Remark}

\newcolumntype{x}[1]{>{\centering\arraybackslash}p{#1}}

\usetikzlibrary{shapes,snakes}
\usetikzlibrary{calc}
\usetikzlibrary{patterns}
\usetikzlibrary{decorations.pathmorphing} 

\newcommand{\basestation}[3]{
\coordinate (a) at (#1,#2);
\draw[line width=(#3)*1.5pt,scale=#3] ($(a)+(0, -1.08)$) -- ($(a)+(0, 1)$);
\draw[line width=(#3)*1.5pt,scale=#3] ($(a)+(0,1)$) .. controls ($(a)+(-0.25,-0.5)$) .. ($(a)+(-0.5,-0.9)$);
\draw[line width=(#3)*1.5pt,scale=#3] ($(a)+(0,1)$) .. controls ($(a)+(0.25,-0.5)$) .. ($(a)+(0.5,-0.9)$);
\draw[line width=(#3)*1.5pt,scale=#3] ($(a)+(0, -0.9)$) -- ($(a)+(-0.35, -0.7)$);
\draw[line width=(#3)*1.5pt,scale=#3] ($(a)+(0, -0.9)$) -- ($(a)+(0.35, -0.7)$);
\draw[line width=(#3)*0.75pt,scale=#3] ($(a)+(-0.35, -0.65)$) -- ($(a)+(0, -0.5)$);
\draw[line width=(#3)*0.75pt,scale=#3] ($(a)+(0.35, -0.65)$) -- ($(a)+(0, -0.5)$);
\draw[line width=(#3)*1pt,scale=#3] ($(a)+(0, -0.6)$) -- ($(a)+(-0.3, -0.45)$);
\draw[line width=(#3)*1pt,scale=#3] ($(a)+(0, -0.6)$) -- ($(a)+(0.3, -0.45)$);
\draw[line width=(#3)*0.5pt,scale=#3] ($(a)+(-0.3, -0.45)$) -- ($(a)+(0, -0.32)$);
\draw[line width=(#3)*0.5pt,scale=#3] ($(a)+(0.3, -0.45)$) -- ($(a)+(0, -0.32)$);
\draw[line width=(#3)*0.75pt,scale=#3] ($(a)+(0, -0.3)$) -- ($(a)+(-0.22, -0.17)$);
\draw[line width=(#3)*0.75pt,scale=#3] ($(a)+(0, -0.3)$) -- ($(a)+(0.22, -0.17)$);
\draw[line width=(#3)*0.5pt,scale=#3] ($(a)+(-0.22, -0.17)$) -- ($(a)+(0, -0.07)$);
\draw[line width=(#3)*0.5pt,scale=#3] ($(a)+(0.22, -0.17)$) -- ($(a)+(0, -0.07)$);;
\draw[line width=(#3)*0.75pt,scale=#3] (a) -- ($(a)+(-0.18, 0.11)$);
\draw[line width=(#3)*0.75pt,scale=#3] (a) -- ($(a)+(0.18, 0.11)$);
\draw[line width=(#3)*0.5pt,scale=#3] ($(a)+(-0.18, 0.11)$) -- ($(a)+(0,0.2)$);
\draw[line width=(#3)*0.5pt,scale=#3] ($(a)+(0.18, 0.11)$) -- ($(a)+(0,0.2)$);   
\draw[line width=(#3)*0.5pt,scale=#3] ($(a)+(0, 0.3)$) -- ($(a)+(-0.1, 0.37)$);
\draw[line width=(#3)*0.5pt,scale=#3] ($(a)+(0, 0.3)$) -- ($(a)+(0.1, 0.37)$);
\draw[line width=(#3)*0.25pt,scale=#3] ($(a)+(-0.1, 0.37)$) -- ($(a)+(0, 0.43)$);
\draw[line width=(#3)*0.25pt,scale=#3] ($(a)+(0.1, 0.37)$) -- ($(a)+(0, 0.43)$);
\draw[line width=(#3)*0.75pt,scale=#3] ($(a)+(0, 1.2)$) -- ($(a)+(0,1)$);
\draw[fill=white,scale=#3] ($(a)+(0, 1.2)$) circle (0.05cm);
\draw[line width=(#3)*1pt,decorate,decoration=expanding waves,scale=#3] ($(a)+(0, 1.2)$) -- ($(a)+(0, 2.5)$);
}
\newcommand{\basestationRed}[3]{
\coordinate (a) at (#1,#2);
\draw[line width=(#3)*1.5pt,scale=#3,color=red] ($(a)+(0, -1.08)$) -- ($(a)+(0, 1)$);
\draw[line width=(#3)*1.5pt,scale=#3,color=red] ($(a)+(0,1)$) .. controls ($(a)+(-0.25,-0.5)$) .. ($(a)+(-0.5,-0.9)$);
\draw[line width=(#3)*1.5pt,scale=#3,color=red] ($(a)+(0,1)$) .. controls ($(a)+(0.25,-0.5)$) .. ($(a)+(0.5,-0.9)$);
\draw[line width=(#3)*1.5pt,scale=#3,color=red] ($(a)+(0, -0.9)$) -- ($(a)+(-0.35, -0.7)$);
\draw[line width=(#3)*1.5pt,scale=#3,color=red] ($(a)+(0, -0.9)$) -- ($(a)+(0.35, -0.7)$);
\draw[line width=(#3)*0.75pt,scale=#3,color=red] ($(a)+(-0.35, -0.65)$) -- ($(a)+(0, -0.5)$);
\draw[line width=(#3)*0.75pt,scale=#3,color=red] ($(a)+(0.35, -0.65)$) -- ($(a)+(0, -0.5)$);
\draw[line width=(#3)*1pt,scale=#3,color=red] ($(a)+(0, -0.6)$) -- ($(a)+(-0.3, -0.45)$);
\draw[line width=(#3)*1pt,scale=#3,color=red] ($(a)+(0, -0.6)$) -- ($(a)+(0.3, -0.45)$);
\draw[line width=(#3)*0.5pt,scale=#3,color=red] ($(a)+(-0.3, -0.45)$) -- ($(a)+(0, -0.32)$);
\draw[line width=(#3)*0.5pt,scale=#3,color=red] ($(a)+(0.3, -0.45)$) -- ($(a)+(0, -0.32)$);
\draw[line width=(#3)*0.75pt,scale=#3,color=red] ($(a)+(0, -0.3)$) -- ($(a)+(-0.22, -0.17)$);
\draw[line width=(#3)*0.75pt,scale=#3,color=red] ($(a)+(0, -0.3)$) -- ($(a)+(0.22, -0.17)$);
\draw[line width=(#3)*0.5pt,scale=#3,color=red] ($(a)+(-0.22, -0.17)$) -- ($(a)+(0, -0.07)$);
\draw[line width=(#3)*0.5pt,scale=#3,color=red] ($(a)+(0.22, -0.17)$) -- ($(a)+(0, -0.07)$);;
\draw[line width=(#3)*0.75pt,scale=#3,color=red] (a) -- ($(a)+(-0.18, 0.11)$);
\draw[line width=(#3)*0.75pt,scale=#3,color=red] (a) -- ($(a)+(0.18, 0.11)$);
\draw[line width=(#3)*0.5pt,scale=#3,color=red] ($(a)+(-0.18, 0.11)$) -- ($(a)+(0,0.2)$);
\draw[line width=(#3)*0.5pt,scale=#3,color=red] ($(a)+(0.18, 0.11)$) -- ($(a)+(0,0.2)$);   
\draw[line width=(#3)*0.5pt,scale=#3,color=red] ($(a)+(0, 0.3)$) -- ($(a)+(-0.1, 0.37)$);
\draw[line width=(#3)*0.5pt,scale=#3,color=red] ($(a)+(0, 0.3)$) -- ($(a)+(0.1, 0.37)$);
\draw[line width=(#3)*0.25pt,scale=#3,color=red] ($(a)+(-0.1, 0.37)$) -- ($(a)+(0, 0.43)$);
\draw[line width=(#3)*0.25pt,scale=#3,color=red] ($(a)+(0.1, 0.37)$) -- ($(a)+(0, 0.43)$);
\draw[line width=(#3)*0.75pt,scale=#3,color=red] ($(a)+(0, 1.2)$) -- ($(a)+(0,1)$);
\draw[fill=white,scale=#3,color=red] ($(a)+(0, 1.2)$) circle (0.05cm);
\draw[line width=(#3)*1pt,decorate,decoration=expanding waves,scale=#3,color=red] ($(a)+(0, 1.2)$) -- ($(a)+(0, 2.5)$);
}
\newcommand{\iPhone}[3]{
\coordinate (a) at (#1,#2);
\draw [line width=0.25pt,rounded corners=(#3)*1mm,fill=white,scale=(#3)] (a)--($(a)+(0.67,0)$)--($(a)+(0.67,1.381)$)--($(a)+(0,1.381)$)--cycle;
\draw [color=gray,line width=0.25pt,rounded corners=(#3)*0.8mm,fill=white,scale=(#3)] ($(a)+(0.015,0.015)$)--($(a)+(0.655,0.015)$)--($(a)+(0.655,1.366)$)--($(a)+(0.015,1.366)$)--cycle;
\draw [line width=0.25pt,rounded corners=(#3)*0.04mm,scale=(#3)] ($(a)+(0.2875,1.266)$)--($(a)+(0.3825,1.266)$)--($(a)+(0.3825,1.281)$)--($(a)+(0.2875,1.281)$)--cycle;
\draw[line width=0.25pt,scale=#3] ($(a)+(0.335,0.09)$) circle (0.055cm);
\draw[line width=0.25pt,scale=#3] ($(a)+(0.335,0.09)$) circle (0.044cm);
\draw[line width=0.25pt,scale=#3] ($(a)+(0.2275,1.2735)$) circle (0.015cm);
\draw[line width=0.25pt,scale=#3] ($(a)+(0.335,1.32)$) circle (0.01cm);
\draw [fill={rgb:black,1;white,4},line width=0.25pt,scale=(#3)] ($(a)+(0.042475,0.170195)$)--($(a)+(0.042475,0.170195)+(0.58505,0.0)$)--($(a)+(0.042475,0.170195)+(0.58505,1.04061)$)--($(a)+(0.042475,0.170195)+(0.0,1.04061)$)--cycle;
}
\newcommand{\iPhoneRed}[3]{
\coordinate (a) at (#1,#2);
\draw [line width=0.25pt,rounded corners=(#3)*1mm,fill=white,scale=(#3)] (a)--($(a)+(0.67,0)$)--($(a)+(0.67,1.381)$)--($(a)+(0,1.381)$)--cycle;
\draw [color=red,line width=0.25pt,rounded corners=(#3)*0.8mm,fill=white,scale=(#3)] ($(a)+(0.015,0.015)$)--($(a)+(0.655,0.015)$)--($(a)+(0.655,1.366)$)--($(a)+(0.015,1.366)$)--cycle;
\draw [line width=0.25pt,rounded corners=(#3)*0.04mm,scale=(#3)] ($(a)+(0.2875,1.266)$)--($(a)+(0.3825,1.266)$)--($(a)+(0.3825,1.281)$)--($(a)+(0.2875,1.281)$)--cycle;
\draw[line width=0.25pt,scale=#3] ($(a)+(0.335,0.09)$) circle (0.055cm);
\draw[line width=0.25pt,scale=#3] ($(a)+(0.335,0.09)$) circle (0.044cm);
\draw[line width=0.25pt,scale=#3] ($(a)+(0.2275,1.2735)$) circle (0.015cm);
\draw[line width=0.25pt,scale=#3] ($(a)+(0.335,1.32)$) circle (0.01cm);
\draw [fill={rgb:black,1;red,4},line width=0.25pt,scale=(#3)] ($(a)+(0.042475,0.170195)$)--($(a)+(0.042475,0.170195)+(0.58505,0.0)$)--($(a)+(0.042475,0.170195)+(0.58505,1.04061)$)--($(a)+(0.042475,0.170195)+(0.0,1.04061)$)--cycle;
}

\begin{document}

\IEEEoverridecommandlockouts
\title{Improper signaling and symbol extensions: How far can we go with Gaussian P2P codebooks in the interfering MAC with TIN?}
\author{
\IEEEauthorblockN{Ali Kariminezhad, Anas Chaaban, and Aydin Sezgin}\\
\thanks{
A. Kariminezhad and A. Sezgin are with the Institute of Digital Communication Systems, Ruhr-Universit\"at Bochum (RUB), Germany (emails: \{ali.kariminezhad, aydin.sezgin\}@rub.de).

A. Chaaban is with the Computer, Electrical, and Mathematical Sciences and Engineering Division, King Abdullah University of Science and Technology (KAUST), 23955-6900 Thuwal, Saudi-Arabia (email: anas.chaaban@kaust.edu.sa).
}}

\maketitle

\begin{abstract}
Meeting the challenges of 5G demands better exploitation of the available spectrum by allowing multiple parties to share resources. For instance, a secondary unlicensed system can share resources with the cellular uplink of a primary licensed system for an improved spectral efficiency. This induces interference which has to be taken into account when designing such a system. A simple yet robust strategy is treating interference as noise (TIN), which is widely adapted in practice. It is thus important to study the capabilities and limitations of TIN in such scenarios. In this paper, we study this scenario modelled as Multiple Access Channel (MAC) interfered by a Point-to-Point (P2P) channel. Here, we focus on rate maximization and power minimization problems separately. We use improper Gaussian signaling (instead of proper) at the transmitters to increase the design flexibility, which offers the freedom of optimizing the transmit signal pseudo-variance in addition to its variance. Furthermore, we allow correlation among the transmitted signals over orthogonal resource basis (i.e., time or frequency) for the purpose of optimal signaling design over the extended channel. We formulate the rate maximization problem as a semidefinite program, and use semidefinite relaxation (SDR) to obtain a near-optimal solution.
Numerical optimizations show that, by improper Gaussian signaling the achievable rates can be improved upto three times depending on the strength of the interfering links. Furthermore, we observe significant benefits in power consumption by improper Gaussian signaling with symbol extensions compared to the traditional proper Gaussian signaling. Interestingly, by minimizing sum power given the solution of the rate maximization problem improves the energy efficiency significantly.
\end{abstract}

\section{Introduction}
\begin{figure}[t]
\centering
\begin{tikzpicture}
\draw[fill=green!10] (0,0) ellipse (6cm and 3cm);
\basestation{0}{0}{1};
\iPhone{1.5}{1.5}{0.5};
\iPhone{1.5}{-2.5}{0.5};
\iPhone{-1.5}{1}{0.5};
\iPhone{4}{0}{0.5};
\iPhone{-4.5}{1}{0.5};

\iPhoneRed{1}{-0.5}{0.5};
\iPhoneRed{2.5}{-0.5}{0.5};

\draw[fill=yellow!30] (-3,-0.5) ellipse (2cm and 1.5cm);
\basestationRed{-3}{-0.5}{0.6};
\iPhoneRed{-2.5}{-0.5}{0.5};

\end{tikzpicture}
\caption{Multiple access channel interfered by communication in a small cell (in yellow) or by a  Device-to-Device communication (in red).}
\label{fig:PIMAC1}
\end{figure}
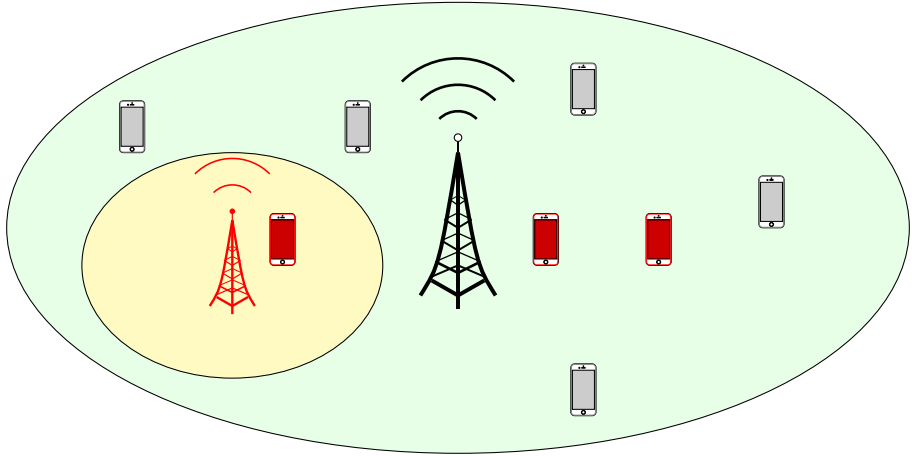

The continuous increase in the demand for high data rates is a challenging issue that confronts today's communication systems. This challenge needs to be addressed in order to enable future systems to cope with this increasing demand. One way to tackle this problem is by allowing shared resources, where multiple users/systems share the same spectrum in order to achieve better performance. By allowing this paradigm of resource sharing, networks become more heterogeneous and more interference-limited. Nevertheless, by allowing shared resources, the performance can be better in comparison to isolating systems by allocating orthogonal resources. This follows since the negative impact of interference can be overcome by the positive impact of increased bandwidth if the transmission is designed properly. As an example, we can think of a primary cellular network sharing resources with another system such as a Device-to-Device (D2D) communication system, a small cell~\cite{Laya2014}, or more generally a secondary cognitive radio~\cite{Dome2012}. Fig. \ref{fig:PIMAC1} depicts a scenario with both D2D nodes and a small cell sharing resources with a cellular network operating in uplink phase. 

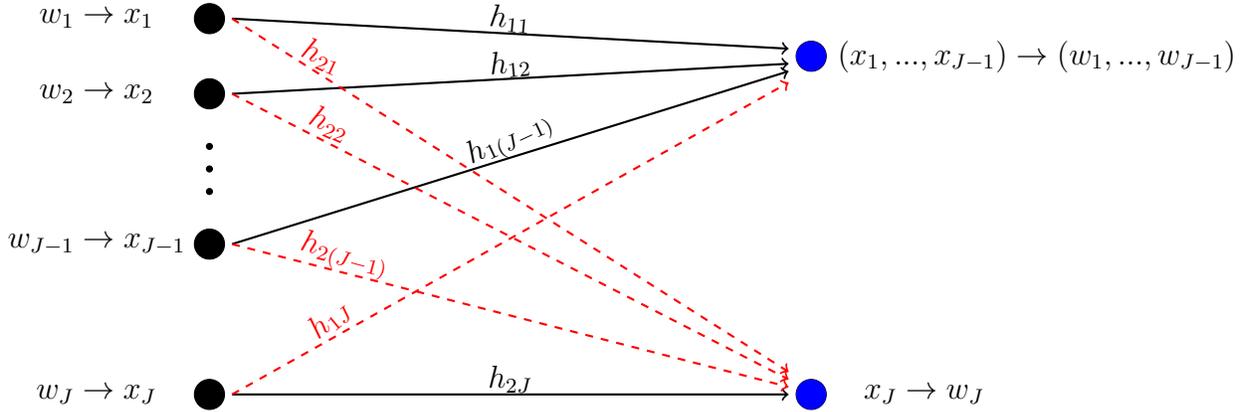
\begin{figure}
\centering
\begin{tikzpicture}
\node at (-1.5,0) {$w_1\rightarrow x_1$};
\node at (-1.5,-1) {$w_2\rightarrow x_2$};
\node at (-1.5,-3) {$w_{J-1}\rightarrow x_{J-1}$};
\node at (-1.5,-5) {$w_J\rightarrow x_J$};

\node at (11,-0.5) {$(x_1, ..., x_{J-1})\rightarrow (w_1, ..., w_{J-1})$};
\node at (9.5,-5) {$x_J\rightarrow w_J$};

\draw (0,0) circle (0.2cm);\fill[black] (0,0) circle (0.2cm);
\draw (0,-1) circle (0.2cm);\fill[black] (0,-1) circle (0.2cm);

\draw (0,-1.7) circle (0.04cm);\fill[black] (0,-1.7) circle (0.04cm);
\draw (0,-2) circle (0.04cm);\fill[black] (0,-2) circle (0.04cm);
\draw (0,-2.3) circle (0.04cm);\fill[black] (0,-2.3) circle (0.04cm);

\draw (0,-3) circle (0.2cm);\fill[black] (0,-3) circle (0.2cm);

\draw (0,-5) circle (0.2cm);\fill[black] (0,-5) circle (0.2cm);

\draw (8,-0.5) circle (0.2cm);\fill[blue] (8,-0.5) circle (0.2cm);
\draw (8,-5) circle (0.2cm);\fill[blue] (8,-5) circle (0.2cm);

\draw[thick,->] (0.3,0) -- (7.7,-0.4);
\draw[thick,->] (0.3,-1) -- (7.7,-0.6);
\draw[thick,->] (0.3,-3) -- (7.7,-0.7);
\draw[thick,->] (0.3,-5) -- (7.7,-5);

\draw[dashed,red,thick,->] (0.3,0) -- (7.7,-4.7);
\draw[dashed,red,thick,->] (0.3,-1) -- (7.7,-4.8);
\draw[dashed,red,thick,->] (0.3,-3) -- (7.7,-4.9);
\draw[dashed,red,thick,->] (0.3,-5) -- (7.7,-0.85);

\node[color=black,rotate=-5] at (4,0) {${ h}_{11}$};
\node[color=black,rotate=5] at (4,-.6) {${ h}_{12}$};
\node[color=black,rotate=25] at (4,-1.6) {${h}_{1(J-1)}$};
\node[color=black,rotate=0] at (4,-4.8) {${ h}_{2J}$};

\node[color=red,rotate=-25] at (1.5,-0.5) {${ h}_{21}$};
\node[color=red,rotate=-25] at (1.6,-1.4) {${ h}_{22}$};
\node[color=red,rotate=-15] at (1.8,-3.15) {${ h}_{2(J-1)}$};
\node[color=red,rotate=40] at (1.6,-4) {${h}_{1J}$};
\end{tikzpicture}
\caption{A Partial Interference Multiple Access Channel (PIMAC). The transmit signals $x_j$ are a function of the messages $w_j$ which are the realizations from the Gaussian codebook.}
\label{fig:PIMAC}
\end{figure}

In this paper, we focus on this aspect in a cellular uplink with shared resources. Namely, we study a network consisting of a MAC sharing its resources with a P2P channel referred to as the partial interfering multiple access channel (PIMAC) as depicted in Fig. \ref{fig:PIMAC},~\cite{chaaban2011interference,chaaban2011capacity,
ChaabanSezgin_EW2011,ZhuShangChenPoor, BuehlerWunder}. As stated earlier, the P2P channel can represent an underlay cognitive system, a pair of D2D communicating devices, or a small cell. Here, the P2P channel is active only if it does not deteriorate the QoS of the primary MAC users \cite{Ying2011,Geir2008,Devroye2006}.

To guarantee good performance, the receivers can employ different interference management strategies. The receivers can either decode interference and subtract it from the received signal to extract the intended signal~\cite{Sato1981,ChaabanSezgin_EW2011}, or simply treat this interference as noise, (TIN)~\cite{Shang2007}. Interestingly, TIN was shown to be optimal for the two-user interference channel (IC) under certain conditions~\cite{Geng2015,Veravalli2009}. Optimality conditions of TIN in the PIMAC were investigated in~\cite{Soheyl2016} where the constant-gap optimality of TIN is studied. We focus on TIN due to its practical simplicity, robustness, and good performance in many practical scenarios.

In this work, we consider a generalized version of TIN which incorporates improper Gaussian signaling~\cite{Schreier2010},~\cite{Schreier2015} instead of the classical proper Gaussian signaling. Compared to proper signaling, improper signaling enables improving the achievable rates of the PIMAC since it enjoys the additional freedom of designing the pseudo-variance in addition to the variance of the transmit signal. For instance, improper signaling was proposed in~\cite{Jafar2009} as a means to improve the Degrees of Freedom (DoF) in the 3-user Interference Channel (IC). In comparison, for the 2-user IC improper signaling does not enhance the DoF, yet improper signaling is useful in the low and moderate SNR (signal to noise power ratio) regime, as it improves the signal-to-interference-plus-noise ratio (SINR) as shown in~\cite{Ho2012,Zeng2013}. In these papers, the authors show that the rate region of the 2-user IC is improved by Gaussian improper signaling compared to Gaussian proper signaling. The PIMAC considered here can be seen as a generalization of the elemental 2-user IC. For instance, using Time Division Multiple Access (TDMA) for the MAC users, the PIMAC can be viewed as a set of separate ICs. Instead of TDMA, in this paper we focus on the general scenario where the users are allowed to simultaneously share the spectrum. The achievable rate tuples of the network are to be determined under this consideration. To this end, we utilize the so-called rate-profile method proposed in \cite{Zhang2010} to characterize the Pareto boundary of the achievable rate region. Herein, the Pareto boundary defines the frontier of the achievable rate region, where an increment in the rate of one user inevitably coincides with a decrement in the rate of at least one of the other users. The problem of characterizing the Pareto boundary by the rate-profile method is non-convex. To overcome this problem, we reformulate the optimization problem as a semidefinite program (SDP) with rank constraints. The reformulated problem is non-convex due to the rank constraints which are then relaxed. The semidefinite relaxation (SDR) is then solved efficiently by interior point methods (i.e., barrier methods~\cite{Boyd2004}). Note that, the optimal solution of SDR may not satisfy the rank constraints of the original problem and we need to determine an approximate solution by the so-called Gaussian randomization process~\cite{Zeng2013}. By numerical evaluation, we demonstrate that under both weak and strong interference, improper signaling improves the Pareto boundary compared to proper signaling. This improvement becomes more apparent when the interference gets stronger.

Characterizing the rate region of a network can not answer all the questions of the network operator. A similar interesting question is to quantify the power requirements under quality-of-service (QoS) constraints. Since the Pareto boundary of the rate region specifies the optimal achievable rate tuples given certain power constraints, there might be several power tuples that provide the same rate tuple on the Pareto boundary. Hence, studying the optimal power allocation for a given rate tuple on the Pareto boundary is of crucial importance. Thus, we consider sum power minimization under QoS constraints, where the QoS is represented by either signal-to-interference-plus-noise ratio (SINR) or rate demands. To proceed with the power minimization problem, we reformulate the complex-valued SISO system model as a real-valued MIMO. Then, we compare the minimum sum power that satisfies the target QoS for various transmission strategies (proper and improper Gaussian signaling) with different complexities. Motivated by cognitive radio networks for optimal resource allocation for the secondary system while satisfying the demands of the primary system, we also investigate the best performance of the P2P channel (considered as a secondary system) when imposing QoS constraints on the MAC (primary system). Having the optimal achievable rate tuples by the relevant optimization problem which characterizes the Pareto boundary and then minimizing the sum power for given achievable rate tuples (a point on the boundary), an energy efficient communication system can be designed. For this purpose, we highlight the benefits of improper Gaussian signaling over an extended symbol compared to proper Gaussian signaling for an energy efficient communication in PIMAC with TIN. Symbol extensions can be realized in the systems with large-enough coherence time, which allows multiple symbols to be precoded jointly. Hence, improper Gaussian signaling over an extended symbol allows correlation between the massages in signal space and the time. For energy efficiency, we address the following questions that need to be applied successively. 
\begin{itemize}
\item How much can the rate region be enlarged for the secondary user by improper signaling while satisfying primary users' QoS demands?
\item How much extra power can be saved by improper signaling over an extended symbol for any given rate tuples on the Pareto boundary? 
\end{itemize}
In this paper, we provide solutions for Pareto boundary characterization and sum power minimization problems separately. Later on, in section \ref{Sec:Num} we discuss the benefits of successive optimization. Furthermore, we propose a joint transmission and reception scheme for a channel with coherence-time several times longer than the symbol duration. For this model, we analyse a strategy based on improper Gaussian signaling, joint beamforming over temporal dimensions (multiple channel uses), successive decoding and TIN. By optimizing the beamformers of the transmitters, we show that improper Gaussian signaling with symbol extensions can save in transmit power while still meeting the QoS constraints.

\subsection{Notation} Throughout the paper, we represent vectors in boldface lower-case letters while the matrices are expressed in boldface upper-case. ${\rm{Tr}}(\bf{A})$, $|{\bf{A}}|$, ${\bf{A}}^{H}$, ${\bf{A}}^{*}$, ${\bf{A}}^{T}$ represent the trace, determinant, hermitian, complex conjugate and transpose of matrix $\mathbf A$, respectively. ${\bf I}_n$ denotes the identity matrix of size $n$. The notation $\otimes$ represents Kronecker product between two matrices.

\subsection{Organization}
The system model and the related assumptions are presented in Sec. \ref{Sec:Model}. In Sec. \ref{Sec:RateMax}, the rate maximization problem is addressed, with which the Pareto boundary of the rate region is determined. Sec. \ref{Sec:PowerMin} considers power minimization under QoS constraints, and with different receiver structures. The results are evaluated numerically in Sec. \ref{Sec:Num}, and we conclude the paper in Sec. \ref{Sec:Conc}.



\section{System Model}
\label{Sec:Model}
The system under investigation, consists of a cellular system operating in the uplink which shares spectrum with a P2P channel. This is modelled as a $J$-user PIMAC, consisting of a MAC with $J-1$ users and a P2P channel. The input-output relation at any given transmission instant can be written as
\begin{align}
y_1&=\underbrace{\sum_{j=1}^{J-1} h_{1j}x_j}_{\text{desired signal}}+\underbrace{h_{1J}x_J+z_1}_{\text{interference + noise}=s_1},\label{eq:y1}\\
y_2&=\underbrace{h_{2J}x_J}_{\text{desired signal}}+\underbrace{\sum_{j=1}^{J-1}h_{2j}x_j
+{z_2}}_{\text{interference + noise}=s_2},\label{eq:y2}
\end{align}
where $h_{ij}$ denotes the complex-valued channel from the $j^{th}$ transmitter to the $i^{th}$ receiver, $z_i$ represents zero-mean additive white Gaussian noise with variance $\sigma^{2}$, i.e., $z_i\sim\mathcal{CN}(0,\sigma^{2})$, $x_j\in\mathbb{C}$ stands for the complex transmit signal from the $j^{th}$ transmitter and $y_i$ is the received signal at the $i^{th}$ receiver. The transmit signals satisfy a power constraint $\mathbb{E}[|X_i|^2]\leq P_i$. We assume that transmitter $i$ encodes an independent message of rate $R_i$, and transmits it over the shared medium. The MAC users communicate with their receiver (a base station (BS)), which receives interference from the P2P channel transmitter, and similarly, the P2P communication observes interference from the MAC users. Note that, the interference-plus-noise terms at the first and second receivers are denoted by $s_1$ and $s_2$, respectively. In what follows, we focus on $J=3$ for clarity in representation, and we comment on the general case ($J>3$) in section \ref{NR2}.

To accomplish this transmission, the following transmission/reception schemes are considered (listed in increasing order of complexity)
\begin{enumerate}[I.]
\item Transmission:
\begin{enumerate}
\item Proper Gaussian signaling,
\item Improper Gaussian signaling, and
\item Improper Gaussian signaling and beamforming over temporal dimensions (symbol extension).
\end{enumerate}
\item Reception:
\begin{enumerate}
\item Scalar-based parallel decoding, 
\item Scalar-based successive decoding (SD),
\item Vector-based SD, and
\item Vector-based SD and time sharing (TS) between decoding orders.
\end{enumerate}
\end{enumerate}
It is important to note here that parallel decoding where the massages are decoded independently and in parallel is the least complex. Decoding the desired massages successively is more complex, but brings rate and power gains in return. 

We assume that each transmitter uses improper Gaussian as input distribution. In order to highlight the performance improvement by using improper Gaussian signaling, we compare different combinations of the transmitter and receiver schemes from rate and power perspectives. We start by studying the problem from a rate maximization perspective.


\section{Rate Maximization}
\label{Sec:RateMax}
Assuming that the receivers treat interference as noise (TIN), and that the MAC receiver uses a MAC-optimal decoding strategy (such as successive decoding combined with time-sharing), we can express the achievable rates of the MAC users as the set of $(R_1,R_2)$ bounded by~\cite{Cover2006}
\begin{align}
R_1\leq I(X_1;Y_1|X_2),\label{eq:r1}\\
R_2\leq I(X_2;Y_1|X_1),\label{eq:r2}\\
R_1+R_2\leq I(X_1,X_2;Y_1),\label{eq:r12}
\end{align}
where $I(X_i;Y_1|X_j)$ is the mutual information between $X_i$ and $Y_1$ given $X_j$, and $I(X_1,X_2;Y_1)$ is the mutual information between $(X_1,X_2)$ and $Y_1$. The third user (P2P user) achieves the following rate by TIN
\begin{align}
R_3\leq I(X_3;Y_2).\label{eq:r3}
\end{align}

Assume that all users in the PIMAC generate their transmit signals from a Gaussian codebook. A Gaussian random variable (RV) is completely characterized by the first-order and second-order moments. The variance of $X$ can be expressed as $C_X=\mathbb{E}[XX^{*}]$, and it completely characterizes the second-order moment of the complex Gaussian RV if and only if it is proper \cite{Schreier2010}. The main idea of improper signaling is to allow non-equal power allocation over the real and imaginary components and allow them to be correlated. The variance in this case does not characterize the second-order moment thoroughly since it does not capture the real-imaginary correlation. Instead, the second-order moment is described by the augmented covariance matrix defined next.

\begin{definition}[~\cite{Schreier2010}] The second-order moment of an improper Gaussian RV $X$ is described by the augmented covariance matrix
\begin{align}
 {\hat{\bf{C}}}_X=
 \begin{bmatrix} 
 C_X & \tilde{C}_X\\
 \tilde{C}^{*}_X & C_X \end{bmatrix},
\end{align}
where, $\tilde{C}_X=\mathrm{E}[XX]$ is the pseudo-variance of $X$.
\end{definition}
By defining the second-order moment of the improper Gaussian random variable, the entropy is defined as,
\begin{definition}[\cite{Schreier2010}] The entropy of an improper Gaussian RV $X$ is
\begin{align}
h(X)=\frac{1}{2}\log({(2\pi{e})^{2}|{\hat{\bf{C}}}_X|})\label{ImpEntropy}.
\end{align}
\end{definition}

The mutual information terms mentioned above can be recast as the subtraction of two entropy terms. Knowing the entropy of an improper Gaussian random variable from \eqref{ImpEntropy}, we can state the following for the P2P user,
\begin{align}
R_3\leq {I({X_3;Y_2})}&=h(Y_2)-h(Y_2|X_3)\nonumber\\
&=\frac{1}{2}\log\frac{|{\hat{\bf{C}}}_{y_2}|}{|{\hat{\bf{C}}}_{s_2}|}\nonumber\\
&=\frac{1}{2}\log\frac{{C^{2}_{y_2}}-{|\tilde{C}_{y_2}|^{2}}}{{C^{2}_{s_2}}-{|\tilde{C}_{s_2}|^{2}}}=L_3,\label{L3}
\end{align}
where $s_2$ is defined in \eqref{eq:y2}. Note that the rate is achievable by improper Gaussian signaling at the transmitter and TIN at the receiver. For the MAC users, the achievable rates using improper signaling can be written similarly as
\begin{align}
R_1\leq\frac{1}{2}\log\frac{{C^{2}_{y_{12}}}-{|\tilde{C}_{y_{12}}|^{2}}}{{C^{2}_{s_1}}-{|\tilde{C}_{s_1}|^{2}}}=L_1,\\
R_2\leq\frac{1}{2}\log\frac{{C^{2}_{y_{11}}}-{|\tilde{C}_{y_{11}}|^{2}}}{{C^{2}_{s_1}}-{|\tilde{C}_{s_1}|^{2}}}=L_2,\\
R_1+R_2\leq\frac{1}{2}\log\frac{{C^{2}_{y_1}}-{|\tilde{C}_{y_1}|^{2}}}{{C^{2}_{s_1}}-{|\tilde{C}_{s_1}|^{2}}}=L_4,\label{extra}
\end{align}
where, $y_{ij}=y_i-h_{ij}x_j$ and $C_{s_1}$,$\hat{C}_{s_1}$, $C_{s_2}$, $\hat{C}_{s_2}$ are the variance and pseudo-variance of the expressions in \eqref{eq:y1}, \eqref{eq:y2}. The variables $L_1$, $L_2$, $L_3$ and $L_4$ are defined for future use in the upcoming optimization problems.

Note that if the P2P user is silent, the system reduces to a Gaussian MAC channel, for which the capacity region can be achieved by proper Gaussian signaling~\cite{Cover2006}. If the P2P user is active however, the achievable rate region of the MAC shrinks. It is interesting to quantify this trade-off between $R_3$ and the set of achievable rates $(R_1,R_2)$. This can be done by studying the Pareto boundary of the achievable rate region.

To characterize the Pareto boundary of the rate region, consider a sum rate $R_{\Sigma}(\boldsymbol{\alpha})$ with ${\boldsymbol{\alpha}}=[\alpha_1,\alpha_2,\alpha_3]\in[0,1]^3$ such that $\sum_{j=1}^3 \alpha_j=1$, so that the users' achievable rates can be expressed as
\begin{align}
R_{j}=\alpha_j{R_{\Sigma}}(\boldsymbol{\alpha}).\label{Rsuma}
\end{align}
The vector $\boldsymbol{\alpha}$ is called the target rate-profile vector. By scanning through feasible rate-profile vectors and maximizing $R_{\Sigma}(\boldsymbol{\alpha})$, we acquire the complete Pareto boundary of the rate region \cite{Zhang2010}.
 
Having defined all the necessary quantities, we can formulate the sum rate maximization problem in a particular scanning direction (i.e., target rate-profile vector) as follows:
\begin{subequations}\label{A1}
\begin{align}
\max_{C_{x_j},\tilde{C}_{x_j},j\in\mathcal{J}} &R_{\Sigma}(\boldsymbol{\alpha}) \tag{\ref{A1}}\\
{\rm{s.t.}} \quad &\alpha_q R_{\Sigma}(\boldsymbol{\alpha})\leq{L_q}, \quad \forall{q\in\mathcal{J}\cup{\{4\}}},\label{merg}   \\
&0\leq{C_{x_j}}\leq{P_{j}}, \quad\forall{j\in\mathcal{J}},\label{pow}\\
&|\tilde{C}_{x_j}|^{2}\leq{C_{x_j}^{2}},\quad\forall{j\in\mathcal{J}},\label{psd}
\end{align}
\end{subequations}
where, $\mathcal{J}=\{1,2,3\}$ is the set of all transmitters and $\alpha_4=\alpha_1+\alpha_2$ is defined in order to fit \eqref{extra} into \eqref{Rsuma}. The variables $L_q,\ \forall q$ are the functions of $C_{x_j},\tilde{C}_{x_j},j\in\mathcal{J}$ and are defined in (\ref{L3})-(\ref{extra}). Transmission power is constrained by \eqref{pow}, and the constraint \eqref{psd} ensures that the augmented covariance matrix is positive semidefinite \cite{Schreier2010}.

Assuming that the optimal value of \eqref{A1} is $R^{*}_{\Sigma}(\boldsymbol{\alpha})$ for a given $\boldsymbol{\alpha}$, the corresponding Pareto-optimal rate tuple is ${\boldsymbol{\alpha}}R^{*}_{\Sigma}$, which is the intersection of the rate region Pareto boundary with the ray in the direction of ${\boldsymbol{\alpha}}$.

Merging the constraints \eqref{merg} into the objective function, problem \eqref{A1} can be expressed as a maximization problem with a weighted Chebyshev objective function~\cite{Ottersten2014},
\begin{subequations}\label{A2}
\begin{align}
\max_{C_{x_j},\tilde{C}_{x_j},j\in\mathcal{J}} &\min_{q\in\mathcal{J}} \frac{L_q}{\alpha_q}\tag{\ref{A2}}\\
{\rm{s.t.}} \quad &0\leq{C_{x_j}}\leq{P_{j}},\quad\forall{j\in\mathcal{J}},\\
&|\tilde{C}_{x_j}|^{2}\leq{C_{x_j}^{2}},\quad\forall{j\in\mathcal{J}}.
\end{align}
\end{subequations}
Problem \eqref{A2} is non-convex. This can be seen by replacing $L_q,\ \forall q$ with the expressions in (\ref{L3})-(\ref{extra}). To obtain a reliable sub-optimal solution of this problem, it can be alternatively written as a SDP with rank constraints by means of some vector definitions. The rank constraints are then relaxed to obtain a relaxed problem (SDR). The solution of the relaxed problem is then projected into the feasible set of the original problem. Details of this procedure are given in the Appendix in order to maintain the reading flow.

The resulting achievable rate region enjoys the benefits of improper signaling, in the form of an enlarged region in comparison with proper signaling. A numerical comparison is given in Section \ref{Sec:Num}.

Besides enlarging the rate region, from a network operator perspective, it is interesting to know the power required to fulfil some QoS requirements. In particular, it is interesting to see if QoS requirements in a cellular uplink e.g., can be met at a lower power even when the spectrum is shared with another system. In the next section, we address this problem for different transmit and receive strategies utilizing improper signaling.

\section{Power Minimization}
\label{Sec:PowerMin}
To formulate the power minimization problem in a compact view, we introduce the real-valued representation of the complex-valued channel model given in (\ref{eq:y1}) and (\ref{eq:y2}). This can be done by stacking the real and imaginary components of the complex transmit symbol $x_j\in\mathbb{C}$ in a vector as ${\bf x}_j=[x_j^{Re} \ x_j^{Im}]^{T}\in\mathbb{R}^{2}$, where $x_j^{Re}$ and $x_j^{Im}$ represent the real and imaginary components of $\mathbf{x}_j$, respectively. Thus, the real-valued equivalent of the system model is as follows:
\begin{align}
{\bf y}_1=\sum_{j=1}^{J-1}{\bf G}_{1j}{\bf x}_j+{\bf G}_{1J}{\bf x}_J+{\bf z}_1,\label{RealRep1}\\
{\bf y}_2={\bf G}_{2J}{\bf x}_J+\sum_{j=1}^{J-1}{\bf G}_{2j}{\bf x}_j+{\bf z}_2,\label{RealRep2}
\end{align}
where ${\bf y}_i\in\mathbb{R}^{2}$, ${\bf x}_j$ and ${\bf z}_i\in\mathbb{R}^{2}$ are the received signal at the $i^{th}$ receiver, transmitted signal from the $j^{th}$ transmitter and the receiver noise, respectively. The channel matrix ${\bf G}_{ij}\in\mathbb{R}^{2\times 2}$ is the real-valued representation of the complex-valued channel which can be expressed as
\begin{align}
{\bf G}_{ij}=
\begin{pmatrix}
h_{ij}^{Re} & -h_{ij}^{Im}
\\
h_{ij}^{Im} & h_{ij}^{Re}
\end{pmatrix}\label{MimoChannelMod}.
\end{align}
By denoting the system model in real domain, the covariance matrix of ${\bf x}_j$ describes its second-order moment thoroughly.

\begin{remark}
In the real representation of the system, the transmit signal covariance matrix captures the power allocation for individual real streams of the signal and their correlation. Thus, by acquiring the freedom for unequal power allocation for the real and imaginary components and potential correlation between them, the covariance matrix of the real representation completely characterizes the second-order moment of a Gaussian random vector.
\end{remark}

With this representation, the complex-valued SISO PIMAC transforms to a real-valued $2\times 2$ MIMO PIMAC. Thus, it is required to find the optimal transmit and receive beamformings. The transmit signal is beamformed  consequently as follows,
\begin{align}
{\bf x}_j=\sum_{k=1}^{2}{{\bf v}_{j_k} d_{j_k}}={\bf V}_j{\bf d}_j,
\end{align}
where $d_{j_1}$ and $d_{j_2}$ are real information streams, ${\bf d}_i=[d_{j1}\ d_{j2}]^T$, and ${\bf v}_{j_1}$ and ${\bf v}_{j_2}$ are the respective beamforming vectors. 

Receiver $i$ applies a receive beamforming matrix ${\bf U}_{i}$ ($2\times 2$ real-valued matrix) to obtain the signal ${\bf\hat y}_{i}$ given by
\begin{align}
{\bf\hat y}_{i}=\sum\limits_{j=1}^{J}{\bf U}_{i}^{T}{\bf G}_{ij}{\bf V}_j{\bf d}_j+{\bf U}_{i}^{T}{\bf n}_i.\label{eq:yHatPower}
\end{align}
The receive beamforming vector of receiver $i\in\mathcal{I}=\{1,2\}$ which corresponds to stream $k\in\mathcal{K}=\{1,2\}$ of transmitter $j\in\mathcal{J}$ is denoted by ${\bf u}_{ijk}$, which is a column of ${\bf U}_{i}$. Note that the MAC receiver is interested in the streams of the two MAC users and needs to decode up to four real streams in total.
\begin{remark}
The optimal variances and pseudo-variances (i.e., $C_{x_j},\ \tilde{C}_{x_j},\ \forall j$) that optimize the system performance in the complex-valued model (\ref{eq:y1}), (\ref{eq:y2}), correspond to a unique transmit beamforming matrices (i.e., ${\bf V}_{j},\ \forall j$) in the real-valued model considering optimal receiver beamforming matrices (i.e., ${\bf U}_{i},\ \forall i$) (\ref{eq:yHatPower}).
\end{remark}
Next, we study the power minimization problem for different reception strategies.

\subsection{Scalar-based Parallel Decoding}
In this section we study a simple receiver that employs single user detection. This means that the $k^{th}$ stream of the $j^{th}$ user is decoded while treating the interference from other streams as noise. Thus, the signal to interference-plus-noise power ratio (SINR) for the $k^{th}$ stream of the $j^{th}$ transmitter at the $i^{th}$ receiver can be written as
\begin{align}
SINR_{ijk}=&\frac{{\bf u}^{T}_{ijk}{\bf T}_{ijk}{\bf u}_{ijk}}{{\bf u}_{ijk}^{T}{\bf F}_{ijk}{\bf u}_{ijk}},\ \forall\{i,j\}\in\mathcal{L},\ \forall k\in\mathcal{K},
\end{align}
where ${\bf T}_{ijk}$ and ${\bf F}_{ijk}$ are the desired and interference-plus-noise covariance matrices, respectively. The set $\mathcal{L}$ is the set of desired receiver-transmitter pairs, i.e., $\mathcal{L}=\left\{\{1,1\},\{1,2\},\{2,3\}
\right\}$. The desired stream covariance matrix is written as
\begin{align}
{\bf T}_{ijk}=&p_{jk}{\bf G}_{ij}{\bf v}_{jk} {\bf v}^{T}_{jk}{\bf G}^{T}_{ij},\ \forall\{i,j\}\in\mathcal{L},\ \forall k\in\mathcal{K},\label{Dijk}
\end{align}
where $p_{jk}$ is the transmit power of the $k^{th}$ real stream of the $j^{th}$ transmitter. 
Given the received signal covariance matrix at receiver $i$ as
\begin{align}
{\bf R}_i= \sum\limits_{l=1}^{3}\sum\limits_{m=1}^{2}p_{lm}{\bf G}_{il}{\bf v}_{lm}{\bf v}^{T}_{lm}{\bf G}^{T}_{il}+\sigma^{2}{\bf I}_2,\quad\forall i\in\mathcal{I},\label{Qi}
\end{align}
we write the interference-plus-noise covariance matrix as,
\begin{align}
{\bf F}_{ijk}=&{\bf R}_i-{\bf T}_{ijk},\ \forall\{i,j\}\in\mathcal{L},\ \forall k\in\mathcal{K},
\end{align}

Our goal is to minimize the transmit power while guaranteeing a certain set of SINR constraints per stream of each user. Hence, the power minimization problem is formulated as follows
\begin{subequations}\label{A3}
\begin{align}
\min_{p_{jk},{\bf v}_{jk},{\bf u}_{ik}} &\sum\limits_{j}\sum\limits_{k}\ p_{jk}
\tag{\ref{A3}}\\
{\rm{s.t.}}\quad &\frac{{\bf u}^{T}_{ijk}{\bf T}_{ijk}{\bf u}_{ijk}}{{\bf u}_{ijk}^{T}{\bf F}_{ijk}{\bf u}_{ijk}}\geq \gamma_{ijk},\ \forall\{i,j\}\in\mathcal{L},\ \forall k\in\mathcal{K},\label{NonConv11}\\
&\sum_{k=1}^{2}p_{jk}\leq P_{j_{max}}, \ \forall j\in\mathcal{J},\label{PowerConsSinr}
\end{align}
\end{subequations}
where $\gamma_{ijk}$ denotes the SINR requirement for the $k^{th}$ stream of the $j^{th}$ transmitter at the $i^{th}$ receiver.

The optimization problem (\ref{A3}) is non-convex due to constraint \eqref{NonConv11}. This constraint, which is the quotient of two convex terms, produces a non-convex set. Due to this non-convexity, we propose two algorithms which provide solutions that outperform state-of-the-art techniques from a minimum power perspective, although possibly sub-optimal. One of these algorithms solves for the beamforming vectors and the transmit power separately, and one which does this jointly. The former has lower complexity, but is expected to be outperformed by the latter. Note that the individual power constraints are imposed by (\ref{PowerConsSinr}). We start with separate optimization.

\subsubsection{Separate Optimization}
A fairly good solution of \eqref{A3} can be obtained by using the following two steps:
\begin{enumerate}[(a)]
\item Optimize the transmit and receive beamforming vectors iteratively for a given transmit power (which fulfils the power constraint), then
\item minimize the transmit power given the sub-optimal beamformers from (a).
\end{enumerate}
Next, we discuss those steps in details.

\paragraph{Beamformer Optimization}
We propose an iterative algorithm which alternates between optimizing the receive beamformers ${\bf u}_{ijk}$ and the transmit beamformers  ${\bf v}_{jk}$ while fixing the transmission power $p_{jk}$.
We start with a given ${\bf v}_{jk}$ (not necessarily optimal), for which we choose ${\bf u}_{ijk}$ that maximizes the SINRs at the respective receivers. For this purpose, we utilize optimal MMSE filters. The MMSE filter for the $k^{th}$ stream of $(i,j)$ pair is written as
\begin{align}
{\bf u}^{*}_{ijk}=
\frac{({\bf F}_{ijk})^{-1}{\bf G}_{ij}{\bf v}_{jk}}{\norm{({\bf F}_{ijk})^{-1}{\bf G}_{ij}{\bf v}_{jk}}},\quad \forall\{i,j\}\in\mathcal{L},\ \forall k\in\mathcal{K},
\label{original1}
\end{align}

\begin{remark} 
Minimum mean-squared error receiver is SINR-optimal among all linear receivers for given transmit beamformer and power (i.e., ${\bf v}_{jk},\ p_{jk},\ \forall j,k$),~\cite{Poor1994}.
\end{remark}

After choosing ${\bf u}_{ijk}$, in order to optimize ${\bf v}_{jk}$, we solve the same problem for the reciprocal network where the roles of the transmitters and receivers are switched. Hence, the SINR-optimal transmit beamformer ${\bf v}_{jk}$ is the MMSE filter corresponding to ${\bf u}_{ijk}$ in the reciprocal system, i.e.,
\begin{align}
\underleftarrow{\bf u}^{*}_{jk}={\bf v}^{*}_{jk}=
\frac{(\underleftarrow{\bf F}_{ijk})^{-1}{\bf G}_{ij}{\bf u}^{*}_{ijk}}{\norm{(\underleftarrow{\bf F}_{ijk})^{-1}{\bf G}_{ij}{\bf u}^{*}_{ijk}}},\quad \forall\{i,j\},k
\label{reciprocal1}
\end{align}
where $\underleftarrow{\bf F}_{ijk}$ is the interference-plus-noise covariance matrix in the reciprocal network.

In the next step, we return to the original network and use the vectors in \eqref{reciprocal1} as the transmit beamformers. Based on those new transmit beamformers, new ${\bf u}_{ijk}$ are computed according to \eqref{original1}. This procedure is repeated iteratively as described in the following algorithm.

\begin{algorithm}
\caption{Beamforming optmization}
\begin{algorithmic}[1]
\State Initialization of ${\bf v}_{jk},\ \forall j,k$ to unit-norm vectors.

\State Calculate ${\bf u}_{ijk}$ according to \eqref{original1} and obtain ${\bf u}^{*}_{ijk}$. 


\State Calculate ${\bf v}_{jk}$ according to \eqref{reciprocal1}.

\State Repeat 2 and 3 until convergence.
\end{algorithmic}
\end{algorithm}

Now that the transmit and receive beamforming vectors have been selected, we turn to the second part of the optimization.

\paragraph{Power Minimization}
Given the transmit and receive beamforming vectors of each user, we minimize the sum power as
\begin{subequations}\label{A4}
\begin{align}
\min_{p_{jk}} &\sum\limits_{j}\sum\limits_{k}\ p_{jk}\tag{\ref{A4}}\\
{\rm{s.t.}}\quad &\frac{{\bf u}^{*^{T}}_{ijk}{\bf T}^{*}_{ijk}(p_{jk}){\bf u}^{*}_{ijk}}{{\bf u}_{ijk}^{*^{T}}{\bf F}^{*}_{ijk}(p_{jk}){\bf u}^{*}_{ijk}}\geq {\gamma_{ijk}},\ \forall i,k\\
&\sum_{k=1}^{2}p_{jk}\leq P_{j_{max}},\ \forall j
\end{align}
\end{subequations}
which is a linear program and can be solved efficiently. Note that the SINR expressions are functions of transmit power $p_{jk}$. We would like to compare this solution with that obtained using joint optimization of beamformers and power (see Sec. \ref{Sec:Num}). Next, we discuss this joint optimization method.

\subsubsection{Joint optimization Problem}
We now consider joint optimization of the beamforming vectors and the transmit power. We embed the transmit power into the transmit beamforming vector, so that is not a unit norm vector anymore. Thus, the beamformer of the $k^{th}$ stream of the $j^{th}$ user is given by
\begin{align}
{\bf v}^{'}_{jk}=\sqrt p_{jk}{\bf v}_{jk}.
\end{align}

By this definition, \eqref{Dijk} and \eqref{Qi} are rewritten as,
\begin{align}
{\bf T}_{ijk}=&{\bf G}_{ij}{\bf Q}_{jk}{\bf G}^{T}_{ij},\ \forall\{i,j\}\in\mathcal{L},\ \forall k\in\mathcal{K},\\
{\bf R}_i=& \sum\limits_{l=1}^{3}\sum\limits_{m=1}^{2}{\bf G}_{il}{\bf Q}_{lm}{\bf G}^{T}_{il}+\sigma^{2}{\bf I}_2,\quad\forall i\in\mathcal{I},
\end{align}
where ${\bf Q}_{jk}={\bf v}^{'}_{jk}{\bf v}^{'^{T}}_{jk}$ is the $k^{th}$ transmit covariance matrix of the $j^{th}$ user. By these definitions, the power minimization problem of \eqref{A3} is recast as
\begin{subequations}\label{A5}
\begin{align}
\min_{{\bf Q}_{jk},{\bf u}_{ik}} &\sum\limits_{j}\sum\limits_{k}{\rm Tr}({\bf Q}_{jk})&
\tag{\ref{A5}}\\
{\rm{s.t.}}\quad &\frac{{\bf u}^{T}_{ijk}{\bf T}_{ijk}{\bf u}_{ijk}}{{\bf u}_{ijk}^{T}{\bf F}_{ijk}{\bf u}_{ijk}}\geq \gamma_{ijk},\ \forall\{i,j\}\in\mathcal{L},\ \forall k\in\mathcal{K},\label{NonConv1}\\
&\sum_{k=1}^{2} {\rm Tr}({\bf Q}_{jk})\leq P_{j_{max}}, \ \forall j\in\mathcal{J},\\
&{\bf Q}_{jk} \succeq 0,\quad\forall{j\in\mathcal{J}\ k\in\mathcal{K}},\\
&{\rm rank}({\bf Q}_{jk})=1,\quad\forall{j\in\mathcal{J}\ k\in\mathcal{K}},\label{rankConsPower}
\end{align}
\end{subequations}  
By dropping the last constraint and fixing the receive beamforming vectors, the relaxed power minimization problem is a SDP which can be solved efficiently. For any given receiver beamforming vector (i.e., ${\bf u}_{ik}$), Problem \eqref{A5} admits rank-1 solutions for the transmit covariance matrices (i.e., ${\bf Q}_{jk}$), even when constraint \eqref{rankConsPower} is dropped~\cite{Huang2010},~\cite{Dahrouj2011}. We solve the problem iteratively based on algorithm 2.

\begin{algorithm}
\caption{Joint beamforming and power optmization}
\begin{algorithmic}[1]
\State Initialize ${\bf v}^{(0)}_{jk},\ \forall j,k$ to unit-norm vectors.

\State Solve \eqref{original1} for ${\bf u}^{*}_{ijk}$.

\State Drop constraint \eqref{rankConsPower} in \eqref{A5}

\State Given ${\bf u}^{*}_{ijk}$ from step 2, solve \eqref{A5}

\If{${\bf Q}^{*}_{jk}$ exists (i.e., \eqref{A5} is feasible), }
\State {Find ${\bf v}^{(1)}_{jk}$ by eigenvalue decomposition of ${\bf Q}^{*}_{jk}$ as
${\bf v}^{(1)}_{jk}=\beta^{\frac{1}{2}}{\bf w}_{jk}$, where $\beta$ and ${\bf w}_{jk}$ are the single eigenvalue and eigenvector of ${\bf Q}^{*}_{jk}$.}
\Else
\State Choose ${\bf v}^{(1)}_{jk}$ from \eqref{reciprocal1}
\EndIf
\State Repeat the procedure from (2)-(9) until convergence.
\end{algorithmic}
\end{algorithm}

This algorithm is also evaluated in Sec \ref{Sec:Num}. Next, we consider another decoding strategy at the receivers.

\subsection{Scalar-based Successive Decoding}
In SD, the receiver removes the contribution of the already decoded signals from the received signal, thus reducing interference in subsequent decoding steps. For instance, given the $(k-1)^{th}$ decoded desired stream of the $j^{th}$ user, i.e., ${\hat x}_{j(k-1)}$, its interference contribution can be cancelled from ${y}_{jk}$. Hence, given the decoded information signals from stream $1$ up to $k-1$, the SINR expression for stream $k$ is written as
\begin{align}
\label{SINRSuccDec}
SINR_{ijk}=&\frac{{\bf u}^{T}_{ijk}{\bf T}_{ijk}{\bf u}_{ijk}}{{\bf u}_{ijk}^{T}{\bf F}^{'}_{ijk}{\bf u}_{ijk}},\ \forall\{i,j\}\in\mathcal{L},\ \forall k\in\mathcal{K},
\end{align}
where
\begin{align}
{\bf F}'_{ijk}=& \sum\limits_{l=1}^{3}\sum\limits_{m=k}^{2}p_{lm}{\bf G}_{il}{\bf v}_{lm}{\bf v}^{T}_{lm}{\bf G}^{T}_{il}+\sigma^{2}{\bf I}_2-{\bf T}_{ijk},\\
=& \sum\limits_{l=1}^{3}\sum\limits_{m=k}^{2}{\bf G}_{il}\mathbf{Q}_{lm}{\bf G}^{T}_{il}+\sigma^{2}{\bf I}_2-{\bf T}_{ijk},\ \forall\{i,j\}, k.
\end{align}
Now, the optimization of the beamformers in this case can be obtained by solving optimization problem \eqref{A5} with the SINR in \eqref{NonConv1} replaced by \eqref{SINRSuccDec}.

Recall that the real representation of the complex SISO channel is equivalent to a $2\times 2$ real MIMO setup \eqref{RealRep1}-\eqref{RealRep2}. In this MIMO setup, each user can transmit 2 real streams, and bounding the rate of each real stream individually is not optimal. Thus, instead of considering the 2 scalar signals separately, we formulate the power minimization problem by considering the 2-dimensional signal vector of users as described next.

\subsection{Vector-based Successive Decoding}

Using the system model in \eqref{RealRep1} and \eqref{RealRep2}, the complex SISO-PIMAC becomes equivalent to a real MIMO-PIMAC, for which the power minimization under rate constraint is a non-convex problem. The rates
\begin{align}
R_1=\frac{1}{2}\log\frac{|\sigma^{2}{\bf I}_2+\sum_{j=1}^{3}{\bf G}_{1j}{\bf Q}_j{\bf G}_{1j}^{T}|}{|\sigma^{2}{\bf I}_2+\sum_{j=2}^{3}{\bf G}_{1j}{\bf Q}_j{\bf G}_{1j}^{T}|},\label{eq:R1}\\
R_2=\frac{1}{2}\log\frac{|\sigma^{2}{\bf I}_2+\sum_{j=2}^{3}{\bf G}_{1j}{\bf Q}_j{\bf G}_{1j}^{T}|}{|\sigma^{2}{\bf I}_2+{\bf G}_{13}{\bf Q}_3{\bf G}_{13}^{T}|},\label{eq:R2}
\end{align}
are achievable for the MAC users by SD (i.e., successive decoding of the first message and then the second message) and TIN,~\cite{Telatar1999}.  The transmit covariance matrix is denoted by ${\bf Q}_j={\bf V}_jE\{{\bf d}_j{\bf d}^{T}_j\}{\bf V}^{T}_j$, where ${\bf V}_j$ and ${\bf d}_j$ are the transmit beamforming matrix and the codeword symbols of the $j^{th}$ user, respectively. Note that $\sigma^{2}$ is the noise variance. For the P2P user, the following rate is achievable~\cite{Telatar1999}
\begin{align}
R_3=\frac{1}{2}\log\frac{|\sigma^{2}{\bf I}_2+\sum_{j=1}^{3}{\bf G}_{2j}{\bf Q}_j{\bf G}_{2j}^{T}|}{|\sigma^{2}{\bf I}_2+\sum_{j=1}^{2}{\bf G}_{2j}{\bf Q}_j{\bf G}_{2j}^{T}|}.
\end{align}

By knowing the achievable rates of the users, the sum power of the network can be minimized guaranteeing certain QoS in terms of achievable rates. Therefore, the power minimization problem can be expressed as
\begin{subequations}\label{A6}
\begin{align}
\min_{{\bf Q}_{j}, j\in\mathcal{J}} &\sum_{j=1}^{3} {\rm Tr}({\bf Q}_j)\tag{\ref{A6}} \\
{\rm{s.t.}} \quad &R_j\geq \beta_j,\quad\forall{j\in\mathcal{J}} \label{RateConsPM1}\\
&{\rm Tr}({\bf Q}_{j})\leq P_{j_{max}}, \ \forall j\in\mathcal{J},\label{b}\\
&{\bf Q}_j \succeq 0,\quad\forall{j\in\mathcal{J}}\label{c}
\end{align}
\end{subequations}
where $\beta_j$ is the $j^{th}$ user rate demand.

This optimization problem is not convex due to the rate constraints \eqref{RateConsPM1}. This can be shown for $j=2$ as
\begin{align}
&R_2\geq \beta_2,\\
&\log\frac{|\sigma^{2}{\bf I}_2+\sum_{j=2}^{3}{\bf G}_{1j}{\bf Q}_j{\bf G}_{1j}^{T}|}{|\sigma^{2}{\bf I}_2+{\bf G}_{13}{\bf Q}_3{\bf G}_{13}^{T}|}\geq 2\beta_2,\\
&\log|\sigma^{2}{\bf I}_2+\sum_{j=2}^{3}{\bf G}_{1j}{\bf Q}_j{\bf G}_{1j}^{T}|-\log|\sigma^{2}{\bf I}_2+{\bf G}_{13}{\bf Q}_3{\bf G}_{13}^{T}|\geq 2\beta_2\label{aa1}.
\end{align}
Since ${\bf Q}_j$ is positive semidefinite, \eqref{aa1} is the difference of two concave functions in the cone of positive semidefinite matrices. This constraint does not produce a convex set intrinsically. In order to get a robust suboptimal solution, we linearise the second term in \eqref{aa1} which yields a convex problem. The linearisation is based on Fenchel's inequality for concave functions. From the Fenchel's inequality, we can express the following lemma, \cite{Boyd2004}.

\begin{lemma} 
For given ${\bf A},{\bf B}\in\mathbb{R}^{a\times b}$, the function $\log|{\bf AXA}^{T}+{\bf BYB}^{T}+{\bf I}_a|$ is upper-bounded by a linear function in $\bf X$, $\bf Y$ as
\begin{align}
&\log|{\bf AXA}^{T}+{\bf BYB}^{T}+{\bf I}_a|\leq\nonumber\\
&\log|{\bf\Gamma}|+{\rm Tr}({\bf \Gamma}^{-1}({\bf AXA}^{T}+{\bf BYB}^{T}+{\bf I}_a))-{\rm Tr}({\bf I}_a),\label{lemma1}
\end{align}
for all ${\bf \Gamma} \in\mathbb{R}^{a\times a}$ so that $\bf\Gamma\succeq 0$. Equality holds when ${\bf \Gamma}={\bf AXA}^{T}+{\bf BYB}^{T}+{\bf I}_a$.
\end{lemma}

Using Lemma 1, we linearise the second term in the rate constraint in \eqref{aa1} and the problem becomes convex. Hence the left hand side of \eqref{aa1} can be directly written as,
\begin{align}
&\log|\sigma^{2}{\bf I}_2+\sum_{j=2}^{3}{\bf G}_{1j}{\bf Q}_j{\bf G}_{1j}^{T}|-\log|\sigma^{2}{\bf I}_2+{\bf G}_{13}{\bf Q}_3{\bf G}_{13}^{T}|\nonumber\\
&\geq\log|\sigma^{2}{\bf I}_2+\sum_{j=2}^{3}{\bf G}_{1j}{\bf Q}_j{\bf G}_{1j}^{T}|-\log|\mathbf{\Gamma}_2|+{\rm Tr}({\mathbf{\Gamma}_2}^{-1}(\sigma^{2}{\bf I}_2+{\bf G}_{13}{\bf Q}_3{\bf G}_{13}^{T}))-{\rm Tr}({\bf I}_2),\label{aaa1}
\end{align}
for all ${\bf \Gamma}_2 \in\mathbb{R}^{2\times 2}$.
The other rate expressions in \eqref{RateConsPM1} can be upper-bounded analogously. Therefore by using Lemma 1, the constraint \eqref{RateConsPM1} turns into the difference of a concave function and a linear function, which is concave. Hence, the problem can be solved efficiently.
\begin{remark}
Notice that for high rate demands, i.e., $\beta_j,\ \forall j\in\mathcal{J}$ and random initializations of ${\bf\Gamma}_j,\ \forall j\in\mathcal{J}$, the constraint set might be empty and renders the problem infeasible. The problem can be solved if the initializations of ${\bf\Gamma}_j$ end up with a non-empty interior (if not reinitialization is required) established by the constraints after concave function linearisation by Lemma 1.
\end{remark}
A unit-norm initializations for ${\bf\Gamma}_j$ ends up in a feasible solution if the noise variance is chosen arbitrarily small. Otherwise, several reinitializations are required to make the problem feasible. Thus, problem \eqref{A6} with modified rates (according to \eqref{aaa1}) is solved for a relatively small noise variance than the actual noise variance, i.e.,  by a factor $\gamma\gg 1$. Then the optimal sum power would be $\gamma\sum{\rm Tr}({\bf Q}_j)$.
The resulting convex optimization problem is solved iteratively so that the linear term approaches the concave term within an arbitrarily small $\epsilon$, i.e., 
\begin{align}
&\log|{\mathbf{\Gamma}_2}|+{\rm Tr}({\mathbf{\Gamma}_2}^{-1}(\sigma^{2}{\bf I}_2+{\bf G}_{13}{\bf Q}_3{\bf G}_{13}^{T})-{\rm Tr}({\bf I}_2)\nonumber \\
&= \log|\sigma^{2}{\bf I}_2+{\bf G}_{13}{\bf Q}_3{\bf G}_{13}^{T}|+\epsilon.\label{epsilonLarge}
\end{align}

It is important to note that, the optimal ${\bf\Gamma}_j\ \forall j$, is determined based on the beamforming solutions in each iteration and is used in the next iteration. As an example, for $R_2$ we get,
\begin{align}
{\bf \Gamma}^{(t+1)}_2=\sigma^{2}{\bf I}_2+{\bf G}_{13}{\bf Q}^{(t)}_3{\bf G}_{13}^{T},\label{Gamma1}
\end{align}
where $t$ represents the iteration index. Note that for an initial ${\bf \Gamma}_j$, the solution is suboptimal and iteration over ${\bf \Gamma}_j$ guides to a more accurate solution. Algorithm 3 explains the procedure briefly, and is evaluated in Sec. \ref{Sec:Num}.
\begin{remark}
By utilizing interior point methods, the transmit covariance matrices are optimized (by inner-loop iteration, i.e., iterations of the interior point methods) at each outer-loop iteration i.e., iterations described in Algorithm 3. The quality of the solutions is improved by further outer-loop iterations (i.e., in the outer-loop iterations the non-convex set produced by \eqref{RateConsPM1} is approximated with a convex set by an arbitrary small approximation error ($\epsilon$). This error goes to zero as the number of iterations ($t$) in algorithm 3 goes to infinity).
\end{remark}

\begin{algorithm}
\caption{Power minimization under rate constraint}
\begin{algorithmic}[1]
\State  $\sigma^{2} \gets \sigma^{2}/\gamma,\ \forall \gamma\gg 1$
\State $t=1$
\State ${\bf\Gamma}^{(1)}_j,\ \forall j\gets{\text{ unit-norm random matrices}}$
\State Solve \eqref{A6} for ${\bf Q}^{(1)}_j,\ \forall j$
\State Calculate $\epsilon^{(1)}$ from \eqref{epsilonLarge}
\State Determine the resolution of the solution, e.g. $\epsilon^{*}$
\While{$\epsilon^{(t)}\geq\epsilon^{*}$}
\State $t=t+1$
\State Calculate ${\bf \Gamma}^{(t)}_j$ from \eqref{Gamma1}
\State Solve \eqref{A6} for ${\bf Q}^{(t)}_j,\ \forall j$
\State Calculate $\epsilon^{(t)}$ from \eqref{epsilonLarge}
\EndWhile
\State \textbf{return} $\gamma\sum{\rm Tr}({\bf Q}^{(t)}_j)$
\end{algorithmic}
\end{algorithm}

All the previous schemes consider the optimization of the beamformers and the transmit power on a symbol-by-symbol basis. Next, we introduce the temporal dimension to the optimization by considering joint transmission over multiple channel uses.

\subsection{Vector-based SD and Symbol Extension}
In this framework, an extended symbol is a vector of multiple transmit symbols in the coherence time of the channel. Thereby, we allow correlation not only between the real and imaginary parts of one symbols, but also between real and imaginary parts of symbols within an extended symbol. For such a system, we revise the model in \eqref{RealRep1} and \eqref{RealRep2} to take this symbol extension into account. For a symbol extension of length $N$, the equivalent channel model is represented as
\begin{align}
{\bf S}_{ij}=\mathbf{I}_{N\times N}\otimes {\bf G}_{ij},
\end{align}
where, ${\bf G}_{ij}$ is the real-valued MIMO channel as in \eqref{MimoChannelMod}. 

By defining the extended channel, we solve the power minimization problem under rate constraint. Note that the equivalent real signaling dimension changes by the factor of $N$. Therefore, the dimension of optimization parameters, i.e. ${\bf Q}_{j},\ \forall{j\in\mathcal{J}}$, depends on the extended symbol length. Thus, we formulate this optimization problem in a same way as \eqref{A6} and solve the problem in an iterative way by using algorithm 3. This leads to lower power requirements for achieving the same rate, as we shall see in the next section.

\section{Numerical Results}
\label{Sec:Num}
In this section, we examine the performance of the joint optimization procedure utilized for optimizing the variance and the pseudo-variance of the complex improper Gaussian signals.
We consider two channel realizations in this section defined as
\begin{align*}
{\bf H}&=
\begin{bmatrix}
h_{11} & h_{12} & h_{13} \\
h_{21} & h_{22} & h_{23}
\end{bmatrix},
\end{align*}
Those channels are given by
\begin{align*}
{\bf H}_1&=
\begin{bmatrix}
2.03e^{-i0.68} & 2.1e^{i2.64} & 3.2e^{i1.48} \\
4.7e^{i1.97} & 4.5e^{-i0.66} & 2.85e^{i2.41}
\end{bmatrix},\\
{\bf H}_2&=
\begin{bmatrix}
3.2e^{-i0.72} & 2.3e^{i2.52} & 1.9e^{i1.35} \\
2.8e^{i1.68} & 2.5e^{-i0.76} & 3.4e^{i2.23}
\end{bmatrix}.
\end{align*}
Note that ${\bf H}_1$ corresponds to a channel with strong interference, while ${\bf H}_2$ is a channel realization with weak interference.

\begin{figure*}[ht]
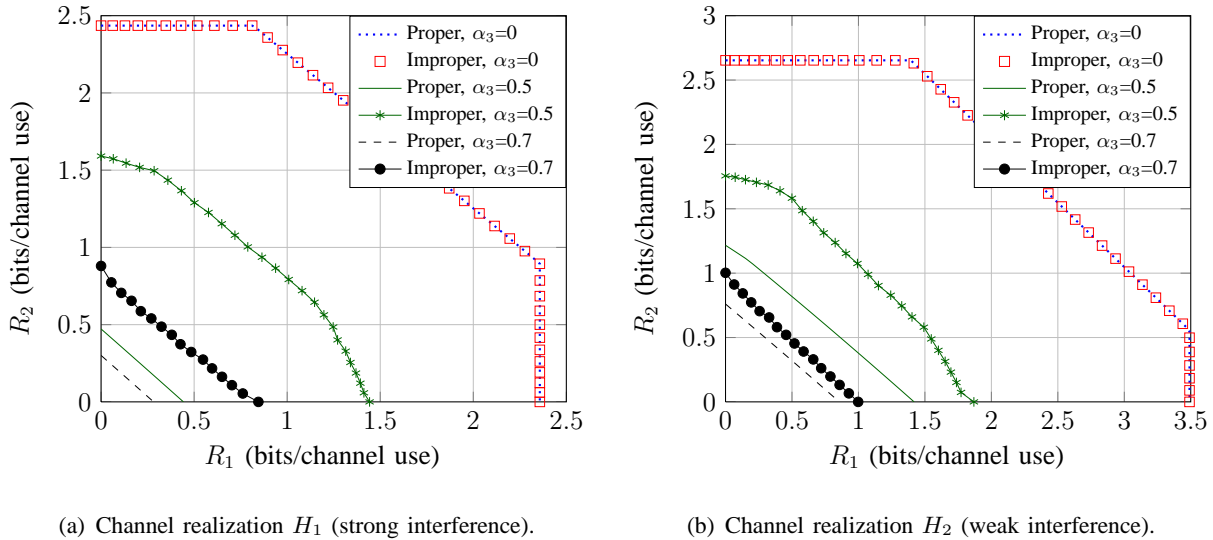

\centering
\subfigure[Channel realization $H_1$ (strong interference).]{
\tikzset{every picture/.style={scale=.95}, every node/.style={scale=.9}}%
\input{RegionR1R2H1}
\label{resulta}
}
\subfigure[Channel realization $H_2$ (weak interference).]{
\tikzset{every picture/.style={scale=.95}, every node/.style={scale=.9}}%
\input{RegionR1R2H2}
\label{resultb}
}
\caption{Comparison of improper and proper signaling at unit maximum transmit power and unit noise variance. The achievable rate region is depicted for the channel realizations $H_1$ and $H_2$ corresponding to strong and weak interference, respectively. $\alpha_3$ is the target rate ratio of the P2P user.}
\label{fig:result}
\end{figure*}

We start by comparing the achievable rate regions using improper signaling in comparison to proper signaling. Fig.~\ref{fig:result} compares those rate regions for the two given channel realizations. According to Fig.~ \ref{fig:result}, in case of silent P2P communication ($\alpha_3=0$ in \eqref{Rsuma}), improper signaling does not enlarge the achievable rate region in comparison to proper signaling. This is due to the fact that proper signaling is optimal in the MAC, which coincides with the PIMAC with $\alpha_3=0$. For active P2P communication, improper signaling outperforms proper signaling from the rate region perspective. The stronger the interference channel, the higher the gain by using improper Gaussian signaling in comparison to proper Gaussian signaling, as shown in Fig.~\ref{fig:result}. According to Fig.~\ref{resulta} which corresponds to high interference, allocating $50\%$ of the sum rate to the P2P communication, improper signaling improves the sum rate of the MAC users at least three times more than proper signaling.

\begin{figure}[ht]
\centering
\tikzset{every picture/.style={scale=.95}, every node/.style={scale=.91}}%
%
%
%
\definecolor{mycolor1}{rgb}{0.00000,0.44706,0.74118}%
\begin{tikzpicture}

\begin{axis}[%
xmin=0,
xmax=2.5,
xtick={  0, 0.5, 1, 1.5, 2,   2.5},
xlabel={$R_1$=$R_2$ (bits/channel use)},
xmajorgrids,
ymin=0,
ymax=4,
ytick={  0, 0.5, 1, 1.5, 2,   2.5, 3 , 3.5, 4},
ylabel={$R_3$ (bits/channel use)},
ymajorgrids,
legend style={at={(axis cs: 2.5,4)},anchor=north east,draw=black,fill=white,legend cell align=left,}
]
\addplot [color=red,solid]
  table[row sep=crcr]{0	3.189\\
0.106298955871207	0.904625\\
0.238995414822772	0.454736842105263\\
0.407580337757025	0.253\\
0.634418498532304	0.128666666666667\\
0.899741224877468	0.065432\\
1.01038597987462	0.046872\\
1.03599402148905	0.0437786666666667\\
1.03838191707311	0.0437786666666667\\
1.08411864722439	0.0392335329341317\\
1.62661330503882	0.000431137724550897\\
};
\addlegendentry{\footnotesize Proper, $H_1$};

\addplot [color=red,solid,mark=*,mark options={solid}]
  table[row sep=crcr]{0	3.18942820309127\\
0.106298955871207	1.91338120568172\\
0.238995414822772	1.91196331858218\\
0.407580337757025	1.90204157619945\\
0.634418498532304	1.90325549559691\\
0.899741224877468	1.79948244975494\\
1.01038597987462	1.34718130649949\\
1.03599402148905	0.887994875562042\\
1.03838191707311	0.519190958536553\\
1.08411864722439	0.240915254938754\\
1.62661330503882	0\\
};
\addlegendentry{\footnotesize Improper, $H_1$};

\addplot [color=blue,dashed]
  table[row sep=crcr]{0	3.65\\
0.0342984941432636	3.3669696969697\\
0.0659466778947817	3.16106896551724\\
0.0966137940401081	2.99658620689655\\
0.126507084935713	2.83777777777778\\
0.157619948845781	2.70166666666667\\
0.189038703659231	2.57338461538462\\
0.220583247814077	2.418\\
0.252167849542017	2.3142\\
0.284622817757683	2.218\\
0.316323684309958	2.1218\\
0.346078787105328	2.03467346938776\\
0.383506224293237	1.92506\\
0.418531251540725	1.84886\\
0.454281877085416	1.75427450980392\\
0.494029478081861	1.66492307692308\\
0.528521991809202	1.60030769230769\\
0.561385796734024	1.52071698113208\\
0.606252879773609	1.44501785714286\\
0.661066918781449	1.34014482097187\\
0.696546133806096	1.29042089552239\\
0.756404735649083	1.19751819248826\\
0.814720193448609	1.11271641791045\\
0.875835314829859	1.03311089010443\\
0.954909063143445	0.936271648351648\\
1.03068130292464	0.846705763397371\\
1.11010772139092	0.740810126582278\\
1.18268907416859	0.6515\\
1.2218066494197	0.602957446808511\\
1.24952860143507	0.579978723404255\\
1.27549722439411	0.546\\
1.29321985145829	0.524\\
1.32069457212417	0.491814814814815\\
1.34412254065423	0.471444444444444\\
1.37280362519859	0.441203389830508\\
1.3848963550964	0.431542372881356\\
1.42620728309759	0.392898305084746\\
1.45709987916808	0.366323076923077\\
1.48178682063596	0.340015384615385\\
1.51242292599379	0.314917808219178\\
1.53546386225223	0.298479452054794\\
1.56517339397385	0.273821917808219\\
1.58346477712628	0.258807228915663\\
1.6227576540851	0.228927710843373\\
1.66091640229943	0.200113402061856\\
1.69288633142494	0.179701030927835\\
1.71961017598553	0.166092783505154\\
1.76676282273076	0.13393670886076\\
1.82152463584289	0.0989999999999998\\
1.90757210189225	0.0559478827361563\\
2.02350740750723	2.22044604925031e-16\\
};
\addlegendentry{\footnotesize Proper, $H_2$};

\addplot [color=blue,dashed,mark=asterisk,mark options={solid}]
  table[row sep=crcr]{0	3.6507604266903\\
0.0659466778947817	3.16544053894952\\
0.126507084935713	2.9096629535214\\
0.189038703659231	2.77256765366872\\
0.252167849542017	2.64776242019118\\
0.316323684309958	2.53058947447966\\
0.383506224293237	2.42887275385717\\
0.454281877085416	2.33630679643928\\
0.528521991809202	2.24621846518911\\
0.606252879773609	2.15556579475061\\
0.696546133806096	2.08963840141829\\
0.814720193448609	2.07383321968737\\
0.954909063143445	2.0689696368108\\
1.11010772139092	2.04942963949092\\
1.2218066494197	1.91998187765952\\
1.27549722439411	1.70066296585881\\
1.32069457212417	1.48578139363969\\
1.37280362519859	1.29205047077514\\
1.42620728309759	1.10927233129813\\
1.48178682063596	0.93586536040166\\
1.53546386225223	0.767731931126113\\
1.58346477712628	0.603224677000489\\
1.66091640229943	0.452977200627116\\
1.71961017598553	0.299062639301831\\
1.82152463584289	0.151793719653574\\
2.02350740750723	0\\
};
\addlegendentry{\footnotesize Improper, $H_2$};

\end{axis}
\end{tikzpicture}%
\caption{Improvement of the achievable rate of the P2P user by improper signaling. Maximum transmit signal power and additive noise variance are set to unity.}
\label{fig:R1R2}
\end{figure}
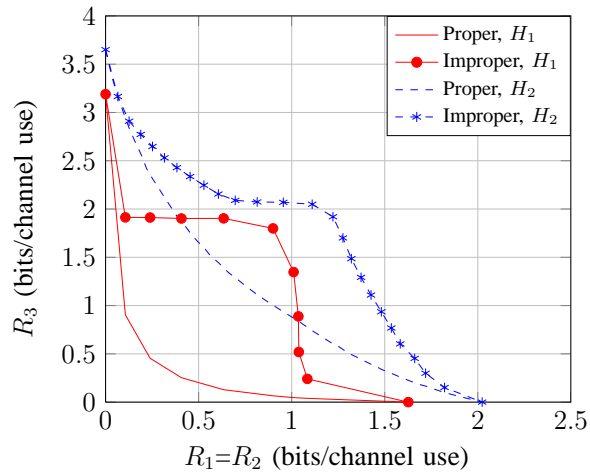

Fig. \ref{fig:R1R2} reflects the gain in $R_3$ achieved by improper signaling compared to proper signaling for equal transmission rate allocation for the users in the MAC. In this scenario, the P2P users can be viewed as an underlay cognitive radio which is activated if the demands of the primary system (MAC) is satisfied. According to this figure, the P2P communication can achieve significantly high data rates by improper signaling. The users in the MAC require $50\%$ of the overall sum rate. By allowing improper signaling, the secondary users (P2P) users can achieve higher rates compared to proper signaling, while maintaining the desired QoS of the MAC users.

\begin{figure*}[ht]
\centering
\subfigure[Channel realization $H_1$ (strong interference).]{
\tikzset{every picture/.style={scale=.95}, every node/.style={scale=.91}}%
\begin{tikzpicture}

\begin{axis}[%
unbounded coords=jump,
xmin=0,
xmax=1.501,
xlabel={SINR demands},
xmajorgrids,
xtick={0,0.25,0.5,0.75,1,1.25,1.5},
ymin=0,
ymax=6,
ylabel={min. $P_\Sigma$ (watts)},
ymajorgrids,
legend style={at={(axis cs: 1.5,0)},anchor= south east, draw=black,fill=white,legend cell align=left}
]
\addplot [color=blue,dashed,mark options={solid}]
  table[row sep=crcr]{0	7.31729214999904e-10\\
0.01	0.00616427423426319\\
0.02	0.0128498254624334\\
0.03	0.0201319901177165\\
0.04	0.0280999506170847\\
0.05	0.0368610698951421\\
0.06	0.0465456695545434\\
0.07	0.0573136162291891\\
0.08	0.0693632863411269\\
0.09	0.082944035099368\\
0.1	0.0983735399882602\\
0.11	0.116063468977234\\
0.12	0.136556718474907\\
0.13	0.160584423218866\\
0.14	0.189155450409801\\
0.15	0.223702104237807\\
0.16	0.266328510278856\\
0.17	0.320256956344723\\
0.18	0.390684521353079\\
0.19	0.486569923739527\\
0.2	0.62478878914428\\
0.21	0.84135965271725\\
0.22	1.22933193448724\\
0.23	2.12562136296464\\
0.24	6.42811661821874\\
};
\addlegendentry{\footnotesize Proper/Par};

\addplot [color=blue, solid,mark options={solid}]
   table[row sep=crcr]{0	7.31729214999904e-10\\
0.01	0.00614019005819528\\
0.02	0.0127470415735976\\
0.03	0.0198841412453652\\
0.04	0.0276254406061737\\
0.05	0.0360584885483328\\
0.06	0.0452876537659316\\
0.07	0.05543850901945\\
0.08	0.0666636277070965\\
0.09	0.079150368218121\\
0.1	0.0931315865086638\\
0.11	0.108900020142876\\
0.12	0.126829874317597\\
0.13	0.147406484896712\\
0.14	0.171271363346111\\
0.15	0.199290693600826\\
0.16	0.232663121176663\\
0.17	0.273096378394071\\
0.18	0.323109788017036\\
0.19	0.386581554558451\\
0.2	0.469807977722492\\
0.21	0.583736284015084\\
0.22	0.749232157671791\\
0.23	1.0116023883635\\
0.24	1.49120138196615\\
0.25	2.64865225378654\\
0.26	9.38630638928601\\
};
\addlegendentry{\footnotesize Proper/Succ};

\addplot [color=black!60!green,mark=o,dashed,mark options={solid}]
  table[row sep=crcr]{0	5.39124568737048e-09\\
0.05	0.221085247785611\\
0.1	0.469364796402521\\
0.15	0.75019801492437\\
0.2	1.07045500785411\\
0.25	1.43909053989328\\
0.3	1.86800149987681\\
0.35	2.37334697798529\\
0.4	2.97765021696333\\
0.45	3.7132847379693\\
0.5	4.62854063963121\\
0.55	5.79881909473152\\
0.6	7.34885772631175\\
0.65	9.50120207191765\\
0.7	12.6960847366356\\
0.75	17.94592515839\\
};
\addlegendentry{\footnotesize Improper/Sep/Par};

\addplot [color=red,dashed,mark=square,mark options={solid}]
  table[row sep=crcr]{0	6.36088664559747e-09\\
0.05	0.0355318575904566\\
0.1	0.0848188982096416\\
0.15	0.155960794599193\\
0.2	0.259117045719167\\
0.25	0.410420892835733\\
0.3	0.628547661562559\\
0.35	0.931341382820118\\
0.4	1.34518365021062\\
0.45	1.90773723264815\\
0.5	2.67175242286637\\
0.55	3.71360439550414\\
0.6	5.14888375289985\\
0.65	7.16264478347453\\
0.7	10.0734478430364\\
0.75	14.4850378137593\\
};
\addlegendentry{\footnotesize Improper/Joint/Par};

\addplot [color=black!60!green,solid,mark=*, mark options={solid}]
  table[row sep=crcr]{0	4.15491158319588e-10\\
0.1	0.130000160692929\\
0.2	0.286568756595372\\
0.3	0.47358177491855\\
0.4	0.695255146259637\\
0.5	0.956145511976018\\
0.6	1.26115102797722\\
0.7	1.61551222177124\\
0.8	2.02481291618467\\
0.9	2.49498130367034\\
1	3.0322908842462\\
1.1	3.64336178051769\\
1.2	4.33516182377904\\
1.3	5.11500789120125\\
1.4	5.99056726034151\\
1.5	6.96985905817958\\
1.6	8.06125577484053\\
1.7	9.27348486740278\\
1.8	10.6156304407127\\
1.9	12.0971350639973\\
2	13.7278013730567\\
};
\addlegendentry{\footnotesize Improper/Sep/Succ};

\addplot [color=red,solid,mark=square*,mark options={solid}]
  table[row sep=crcr]{0	6.36088664308163e-09\\
0.1	0.0762087645849871\\
0.2	0.187989387997936\\
0.3	0.334496232087999\\
0.4	0.517264819677624\\
0.5	0.739784701700755\\
0.6	1.00640937152509\\
0.7	1.32199532792767\\
0.8	1.69177649637274\\
0.9	2.12131433196518\\
1	2.61647559016899\\
1.1	3.18342175983972\\
1.2	3.82860358968386\\
1.3	4.55875826134174\\
1.4	5.38090779196386\\
1.5	6.30235825201481\\
1.6	7.33069933705868\\
1.7	8.4738040577671\\
1.8	9.7398289150005\\
1.9	11.137213892013\\
2	12.674681746684\\
};
\addlegendentry{\footnotesize Improper/Joint/Succ};
\end{axis}
\end{tikzpicture}%
\label{powerMinSinrA}
}
\subfigure[Channel realization $H_2$ (weak interference).]{
\tikzset{every picture/.style={scale=.95}, every node/.style={scale=.91}}%
\begin{tikzpicture}

\begin{axis}[%
unbounded coords=jump,
xmin=0,
xmax=2.501,
xlabel={SINR demands},
xmajorgrids,
ymin=0,
ymax=6,
ylabel={min. $P_\Sigma$ (watts)},
ymajorgrids,
legend style={at={(axis cs: 2.5,0)},anchor=south east, draw=black,fill=white,legend cell align=left}
]
\addplot [color=blue,dashed,mark options={solid}]
  table[row sep=crcr]{0	8.75627824627358e-10\\
0.01	0.00378731154649476\\
0.02	0.00768823537656012\\
0.03	0.0117079256418321\\
0.04	0.0158517574647396\\
0.05	0.0201254030641771\\
0.06	0.0245349126147223\\
0.07	0.0290867112069565\\
0.08	0.0337876361805009\\
0.09	0.0386449772949017\\
0.1	0.0436665060134281\\
0.11	0.0488605267509365\\
0.12	0.0542359080007638\\
0.13	0.0598021449417366\\
0.14	0.0655694097738799\\
0.15	0.0715486036352261\\
0.16	0.0777514634539394\\
0.17	0.0841905811650652\\
0.18	0.0908795255455634\\
0.19	0.0978329278031982\\
0.2	0.105066587573632\\
0.21	0.11259759186701\\
0.22	0.120444448852178\\
0.23	0.128627238678601\\
0.24	0.137167783924752\\
0.25	0.146089842713162\\
0.26	0.155419328093789\\
0.27	0.165184557964692\\
0.28	0.175416540593393\\
0.29	0.186149301810101\\
0.3	0.19742026114342\\
0.31	0.209270665659676\\
0.32	0.221746092218365\\
0.33	0.234897031357773\\
0.34	0.248779565429171\\
0.35	0.263456167103502\\
0.36	0.278996635623551\\
0.37	0.295479202333582\\
0.38	0.312991844779281\\
0.39	0.331633852516689\\
0.4	0.35151770548346\\
0.41	0.372771339504025\\
0.42	0.395540894627652\\
0.43	0.41999407394321\\
0.44	0.446324271601028\\
0.45	0.474755695258478\\
0.46	0.505549732185458\\
0.47	0.539013027495661\\
0.48	0.575507693833436\\
0.49	0.615464456025396\\
0.5	0.659399709010917\\
0.51	0.707937931217098\\
0.52	0.761841523016399\\
0.53	0.822051110535897\\
0.54	0.889740856360005\\
0.55	0.966395733177493\\
0.56	1.05392165303372\\
0.57	1.15480599480908\\
0.58	1.27235770457425\\
0.59	1.41107679848527\\
0.6	1.57724374456188\\
0.61	1.77989618701055\\
0.62	2.03252872997723\\
0.63	2.35622676809152\\
0.64	2.78587715317914\\
0.65	3.38366981561261\\
0.66	4.27230455508217\\
0.67	5.73224077904043\\
0.68	8.57534811362327\\
0.69	16.536367346102\\
};
\addlegendentry{\footnotesize Proper/Par};

\addplot [color=blue, solid,mark options={solid}]
   table[row sep=crcr]{0	8.75627824627358e-10\\
0.01	0.00376794561396812\\
0.02	0.00760889406363421\\
0.03	0.0115250118604066\\
0.04	0.0155184337336954\\
0.05	0.0195913907315924\\
0.06	0.0237461867623432\\
0.07	0.0279852222472442\\
0.08	0.0323109936853069\\
0.09	0.0367261067616485\\
0.1	0.0412332664656252\\
0.11	0.0458352981952568\\
0.12	0.0505351448371564\\
0.13	0.0553358749825364\\
0.14	0.0602406895017642\\
0.15	0.0652529273456554\\
0.16	0.070376075088387\\
0.17	0.0756137742276177\\
0.18	0.0809698298141011\\
0.19	0.0864482196692045\\
0.2	0.0920531042387272\\
0.21	0.0977888371268149\\
0.22	0.103659976525242\\
0.23	0.109671297633659\\
0.24	0.115827803625929\\
0.25	0.12213474183439\\
0.26	0.128597617841263\\
0.27	0.135222207727333\\
0.28	0.142014590707757\\
0.29	0.148981144112155\\
0.3	0.156128579352068\\
0.31	0.163463959410131\\
0.32	0.17099472209895\\
0.33	0.178728705239741\\
0.34	0.186674173881695\\
0.35	0.194839849366772\\
0.36	0.203234941696634\\
0.37	0.211869183979203\\
0.38	0.220752870043638\\
0.39	0.229896895362741\\
0.4	0.239312801554968\\
0.41	0.249012824954825\\
0.42	0.259009949618404\\
0.43	0.269317965256103\\
0.44	0.279951530682459\\
0.45	0.290926243358123\\
0.46	0.302258715206926\\
0.47	0.313966657029593\\
0.48	0.326068970568382\\
0.49	0.338585854527813\\
0.5	0.351538900013251\\
0.51	0.364951240199488\\
0.52	0.378847667655629\\
0.53	0.39325480509662\\
0.54	0.408201268044075\\
0.55	0.42371782970632\\
0.56	0.439837667585589\\
0.57	0.456596586372676\\
0.58	0.474033283680734\\
0.59	0.49218964816438\\
0.6	0.511111095487704\\
0.61	0.530846947159429\\
0.62	0.551450860179652\\
0.63	0.572981313829917\\
0.64	0.595502163351234\\
0.65	0.619083272442662\\
0.66	0.643801236909816\\
0.67	0.66974021548776\\
0.68	0.696992886815202\\
0.69	0.72566155548665\\
0.7	0.755859435038761\\
0.71	0.787712142012575\\
0.72	0.821359441483625\\
0.73	0.856957296823457\\
0.74	0.894680287000043\\
0.75	0.934724468863625\\
0.76	0.977310784809814\\
0.77	1.02268914053169\\
0.78	1.07114331138176\\
0.79	1.12299688185878\\
0.8	1.17862047975157\\
0.81	1.23844064748799\\
0.82	1.30295079969673\\
0.83	1.3727248627454\\
0.84	1.44843446352591\\
0.85	1.53087032630044\\
0.86	1.6209704468925\\
0.87	1.71985564173337\\
0.88	1.82887642296639\\
0.89	1.94967511495774\\
0.9	2.08426912968359\\
0.91	2.23516453269083\\
0.92	2.40551351836939\\
0.93	2.59933705269332\\
0.94	2.82184619506677\\
0.95	3.07991768637744\\
0.96	3.38281724341974\\
0.97	3.7433356439419\\
0.98	4.17964126612991\\
0.99	4.71843814078681\\
1	5.40064569606448\\
1.01	6.29230998100477\\
1.02	7.50738757527473\\
1.03	9.26081304890123\\
1.04	12.0121737838372\\
1.05	16.9522683562093\\
};
\addlegendentry{\footnotesize Proper/Succ};

\addplot [color=black!60!green,dashed,mark=o,mark options={solid}]
  table[row sep=crcr]{0	3.40027641192851e-09\\
0.05	0.021622424922408\\
0.1	0.0466414959005202\\
0.15	0.0759122865788413\\
0.2	0.11060380344775\\
0.25	0.152357496046906\\
0.3	0.203552634152525\\
0.35	0.26777324449766\\
0.4	0.350681085034303\\
0.45	0.46177640779745\\
0.5	0.618314799217553\\
0.55	0.855250008530253\\
0.6	1.25574877029045\\
0.65	2.07807165087594\\
0.7	4.72839455992537\\
0.73	14.8023164090719\\
};
\addlegendentry{\footnotesize Improper/Sep/Par};

\addplot [color=red,dashed,mark=square,mark options={solid}]
  table[row sep=crcr]{0	4.79183615447403e-09\\
0.05	0.0200250089213391\\
0.1	0.0431594839764241\\
0.15	0.0701787711094956\\
0.2	0.102086625056792\\
0.25	0.140258609078364\\
0.3	0.186611178468156\\
0.35	0.243899241518297\\
0.4	0.316203066296313\\
0.45	0.40963632303579\\
0.5	0.53349623867874\\
0.55	0.701768584197542\\
0.6	0.934429070020166\\
0.65	1.26215914204258\\
0.7	1.74836989437327\\
0.75	2.5086368902133\\
0.8	3.78709815939592\\
0.85	6.19777836791229\\
0.9	11.8264072148596\\
};
\addlegendentry{\footnotesize Improper/Joint/Par};

\addplot [color=black!60!green,solid,mark=*, mark options={solid}]
table[row sep=crcr]{0	4.60846966388523e-10\\
0.1	0.11863869695832\\
0.2	0.249522078747611\\
0.3	0.392911045191598\\
0.4	0.549090572005353\\
0.5	0.718370085878294\\
0.6	0.901083892291882\\
0.7	1.09759164330337\\
0.8	1.30827886792238\\
0.9	1.53355755351733\\
1	1.7738668089046\\
1.1	2.02967347362142\\
1.2	2.30147308397571\\
1.3	2.58979054518948\\
1.4	2.89518119921372\\
1.5	3.21823170218117\\
1.6	3.55956140831032\\
1.7	3.91982333867567\\
1.8	4.29970571981768\\
1.9	4.69993328992381\\
2	5.12126909889043\\
2.1	5.56451619415559\\
2.2	6.03051944267724\\
2.3	6.52016782705729\\
2.4	7.03439657989037\\
2.5	7.57418980870468\\
2.6	8.14058311894138\\
2.7	8.73466665072765\\
2.8	9.35758832598473\\
2.9	10.0105574022475\\
3	10.6948483545026\\
};
\addlegendentry{\footnotesize Improper/Sep/Succ};

\addplot [color=red,solid,mark=square*,mark options={solid}]
table[row sep=crcr]{0	4.7918361542721e-09\\
0.1	0.0412336831818073\\
0.2	0.0920530713530508\\
0.3	0.156128523158029\\
0.4	0.239312721502906\\
0.5	0.340648115785988\\
0.6	0.384880541782401\\
0.7	0.49057478718647\\
0.8	0.611569927772193\\
0.9	0.749075830963994\\
1	0.904218573734146\\
1.1	1.07802728753587\\
1.2	1.27143260901247\\
1.3	1.48527599336045\\
1.4	1.72032663410682\\
1.5	1.97730178737766\\
1.6	2.25688582249865\\
1.7	2.55974657018536\\
1.8	2.8865474306501\\
1.9	3.23795597852691\\
2	3.61464912307324\\
2.1	4.01731619901564\\
2.2	4.4466604457643\\
2.3	4.90339921811121\\
2.4	5.38826359424431\\
2.5	5.90199757255799\\
2.6	6.44535703402621\\
2.7	7.01910866983665\\
2.8	7.62402884944702\\
2.9	8.26090260600484\\
3	8.9305227733069\\
3.1	9.63368892315116\\
3.2	10.3712066952587\\
3.3	11.1438872387592\\
3.4	11.9525463408281\\
3.5	12.7980041589765\\
};    
\addlegendentry{\footnotesize Improper/Joint/Succ};
\end{axis}
\end{tikzpicture}%
\label{powerMinSinrB}
}
\caption{Minimum power required to fulfil certain SINR constraints which are equal for all streams of all users. Proper and improper signaling are compared. Note that parallel decoding and successive decoding are denoted by "Par" and "Succ", respectively.}
\label{PowerMinSinr}
\end{figure*}
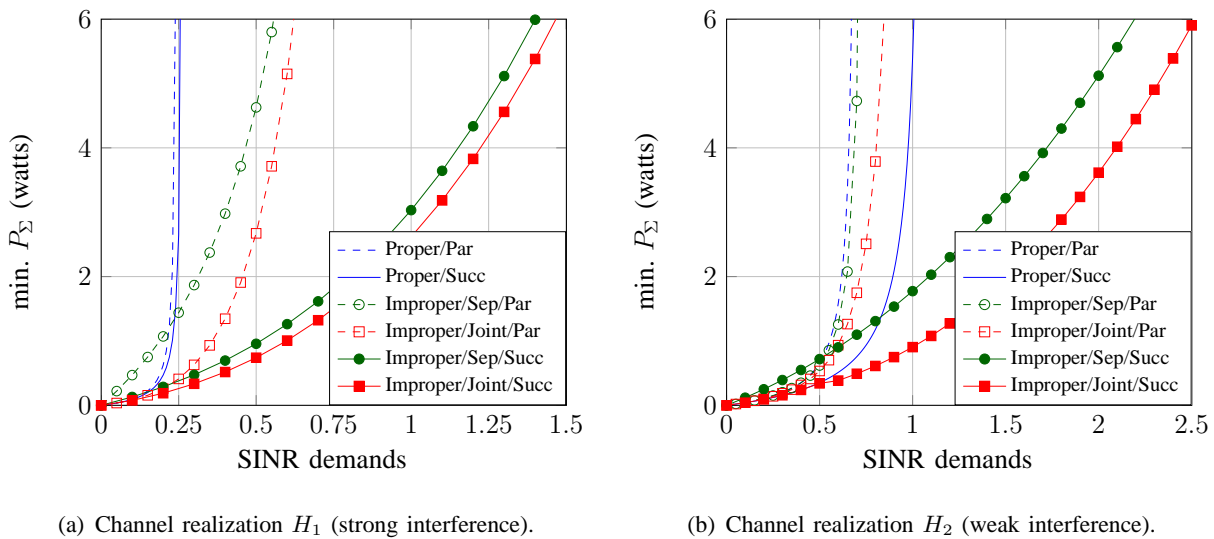

In Fig. \ref{PowerMinSinr}, we illustrate the minimum required power for achieving certain SINRs per stream per user in the PIMAC. It is shown that by allowing improper signaling, the same SINR can be achieved by less power consumption. Furthermore, receivers which are capable of SD perform better. According to Fig.~\ref{powerMinSinrA}, at strong interference, increasing transmit complexity (beamforming) by improper signaling performs is more efficient than increasing decoding complexity by SD from the power perspective. Recall that power minimization for proper Gaussian signaling is a convex problem which can be solved efficiently, but power minimization problem for improper Gaussian signaling suffers from non-convexity. Nevertheless, the proposed algorithm finds a reliable suboptimal beamforming and power solution with which certain QoS demands can be fulfilled with less power compared to proper Gaussian signaling.

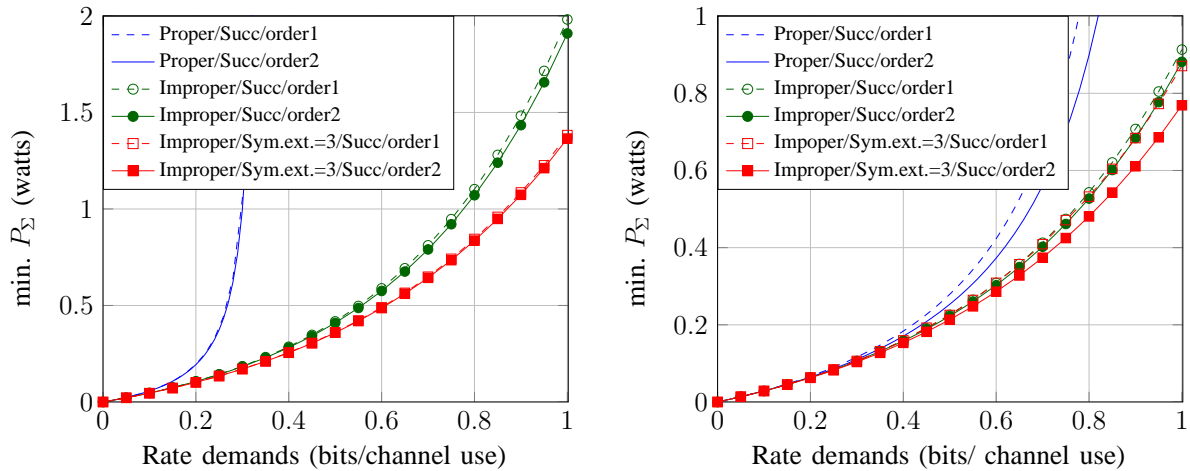
\begin{figure*}
\centering
\subfigure[Channel realization $H_1$ (strong interference).]{
\tikzset{every picture/.style={scale=.95}, every node/.style={scale=.91}}%
\begin{tikzpicture}

\begin{axis}[%
unbounded coords=jump,
xmin=0,
xmax=1.002,
xlabel={Rate demands (bits/channel use)},
xmajorgrids,
ymin=0,
ymax=2.001,
ylabel={min. $P_\Sigma$ (watts)},
ymajorgrids,
legend style={at={(axis cs: 0, 2.001)},anchor=north west, draw=black,fill=white,legend cell align=left}
]
\addplot [color=blue,dashed]
  table[row sep=crcr]{0	3.49700204656333e-05\\
0.02	0.00869806922906024\\
0.04	0.0184951776295835\\
0.06	0.0296752807049882\\
0.08	0.0424538818191233\\
0.1	0.057349922470488\\
0.12	0.0749236867119012\\
0.14	0.0959889479707766\\
0.16	0.121728319752411\\
0.18	0.153919376388079\\
0.2	0.195371055439157\\
0.22	0.250792910490256\\
0.24	0.328730344474492\\
0.26	0.446481960572882\\
0.28	0.64509846253673\\
0.3	1.05177000048688\\
0.32	2.35366433149143\\
};
\addlegendentry{\footnotesize Proper/Succ/order1};

\addplot [color=blue,solid]
  table[row sep=crcr]{0	0\\
0.01	 0.00427423940800073\\
0.02	 0.00869475946403186\\
0.03	 0.0134372355874883\\
0.04	 0.0184796255733925\\
0.05	 0.0239269996734735\\
0.06	 0.0296421474630702\\
0.07	 0.0357740666559294\\
0.08	 0.0423725309440058\\
0.09	 0.0494921465736047\\
0.1	 0.0571947624808415\\
0.11	 0.0655536430625144\\
0.12	 0.0746550247950405\\
0.13	 0.0846095685742583\\
0.14	 0.095543197962716\\
0.15 	0.107611289230579\\
0.16	 0.121002117322078\\
0.17	 0.135948897741732\\
0.18 	0.15274283562644\\
0.19	 0.171752399623599\\
0.2	0.193450476962458\\
0.21	 0.218454886399822\\
0.22	 0.247587917382849\\
0.23	 0.281968427853753\\
0.24	 0.323159727212767\\
0.25	 0.373442969107063\\
0.26	 0.436107325903577\\
0.27	 0.516490953535545\\
0.28	 0.62332747851852\\
0.29	 0.772299035637283\\
0.3	 0.994234701075905\\
0.31	 1.36056853186268\\
0.32	 2.07995272664991\\
};
\addlegendentry{\footnotesize Proper/Succ/order2};

\addplot [color=black!60!green,dashed,mark=o,mark options={solid}]
  table[row sep=crcr]{0	0\\
0.05	0.0219448902251515\\
0.1	0.0467838783157341\\
0.15	0.0750575395583147\\
0.2	0.107061668818482\\
0.25	0.143600812815765\\
0.3	0.18506936714431\\
0.35	0.232404044486037\\
0.4	0.286217307545056\\
0.45	0.347616821896355\\
0.5	0.417646443127643\\
0.55	0.497499150944562\\
0.6	0.588678216479609\\
0.65	0.692734035207266\\
0.7	0.811622432324267\\
0.75	0.947410338512845\\
0.8	1.10260078037394\\
0.85	1.28012913619964\\
0.9	1.48304301585175\\
0.95	1.71523116893231\\
1	1.98192323755895\\
};
\addlegendentry{\footnotesize Improper/Succ/order1};

\addplot [color=black!60!green,solid,mark=*]
  table[row sep=crcr]{0	0\\
0.05	0.0218619605126755\\
0.1	0.0466013619230065\\
0.15	0.0745464640032028\\
0.2	0.106246405590303\\
0.25	0.142186811529334\\
0.3	0.182953604148553\\
0.35	0.22927708175996\\
0.4	0.281845708576818\\
0.45	0.341643159235335\\
0.5	0.409664887455691\\
0.55	0.487094947960282\\
0.6	0.575260530731479\\
0.65	0.675712353533867\\
0.7	0.790205905111655\\
0.75	0.920821875218304\\
0.8	1.07110233856216\\
0.85	1.23983503756143\\
0.9	1.43404907411068\\
0.95	1.65619236632629\\
1	1.90941897252454\\
};
\addlegendentry{\footnotesize Improper/Succ/order2};

\addplot [color=red,dashed,mark=square,mark options={solid}]
  table[row sep=crcr]{0	0\\
0.05	0.0216934526326547\\
0.1	0.0456920253053325\\
0.15	0.0723142190081568\\
0.2	0.101844812732144\\
0.25	0.134624536861946\\
0.3	0.171005084725884\\
0.35	0.211369960497943\\
0.4	0.256234802223403\\
0.45	0.305863833603525\\
0.5	0.361024030613139\\
0.55	0.422210891787619\\
0.6	0.490114867334646\\
0.65	0.565435302383678\\
0.7	0.649033987511943\\
0.75	0.741778588827696\\
0.8	0.844662375078138\\
0.85	0.958884705603225\\
0.9	1.08552392528574\\
0.95	1.22610441385879\\
1	1.38263211128043\\
};
\addlegendentry{\footnotesize Improper/Sym.ext.=3/Succ/order1};

\addplot [color=red,solid,mark=square*,mark options={solid}]
  table[row sep=crcr]{0	0\\
0.05	0.0216600091950091\\
0.1	0.0456089168506755\\
0.15	0.0721943970746109\\
0.2	0.101601878708492\\
0.25	0.13426318933447\\
0.3	0.170364657219103\\
0.35	0.2105497812585\\
0.4	0.254857343063952\\
0.45	0.304170351109241\\
0.5	0.358694149161087\\
0.55	0.419214668519316\\
0.6	0.486413496673388\\
0.65	0.560705377974931\\
0.7	0.64320557375647\\
0.75	0.734679485668977\\
0.8	0.836074852500022\\
0.85	0.948528447115415\\
0.9	1.0732230900463\\
0.95	1.21155194562621\\
1	1.3648826365297\\
};
\addlegendentry{\footnotesize Improper/Sym.ext.=3/Succ/order2};

\end{axis}
\end{tikzpicture}%
\label{PowerMinRateA}
}
\subfigure[Channel realization $H_2$ (weak interference).]{
\tikzset{every picture/.style={scale=.95}, every node/.style={scale=.91}}%
\begin{tikzpicture}

\begin{axis}[%
unbounded coords=jump,
xmin=0,
xmax=1.002,
xlabel={Rate demands (bits/ channel use)},
xmajorgrids,
ymin=0,
ymax=1.001,
ylabel={min. $P_\Sigma$ (watts)},
ymajorgrids,
legend style={at={(axis cs: 0,1.001)},anchor=north west,draw=black,fill=white,legend cell align=left}
]
\addplot [color=blue,dashed]
  table[row sep=crcr]{0	2.73238768324168e-05\\
0.01	0.00266153161706175\\
0.02	0.00537111023198891\\
0.03	0.00804392841226732\\
0.04	0.0108553442627819\\
0.05	0.0137353352413257\\
0.06	0.0167558924801343\\
0.07	0.0197642540251624\\
0.08	0.0228488617775206\\
0.09	0.0260138142997178\\
0.1	0.0292617984523372\\
0.11	0.0325953355695704\\
0.12	0.0360182232425912\\
0.13	0.0395325374622731\\
0.14	0.0431381858284788\\
0.15	0.0468397766194313\\
0.16	0.0506401673113838\\
0.17	0.054541917552036\\
0.18	0.058548818990163\\
0.19	0.0626602568612843\\
0.2	0.0668895356880149\\
0.21	0.0712362109458574\\
0.22	0.0757047028306118\\
0.23	0.0802996741229877\\
0.24	0.0850260311313332\\
0.25	0.0898888679837803\\
0.26	0.0948935375299732\\
0.27	0.100045848247293\\
0.28	0.105351825800416\\
0.29	0.110817832721887\\
0.3	0.116450690755186\\
0.31	0.122257487694368\\
0.32	0.128245680046751\\
0.33	0.134423219756742\\
0.34	0.140798584046834\\
0.35	0.147380738420958\\
0.36	0.154179240202661\\
0.37	0.161204353326317\\
0.38	0.168466659517817\\
0.39	0.175977607480849\\
0.4	0.183749387199853\\
0.41	0.191795008091596\\
0.42	0.2001283842171\\
0.43	0.208764381296961\\
0.44	0.21771891318136\\
0.45	0.227009036070554\\
0.46	0.236653053036275\\
0.47	0.246670628502103\\
0.48	0.257082920620921\\
0.49	0.267912680588496\\
0.5	0.27918455324272\\
0.51	0.290925111260851\\
0.52	0.303163109982806\\
0.53	0.315929715831356\\
0.54	0.329258764686465\\
0.55	0.343187049862804\\
0.56	0.357754657959982\\
0.57	0.373005343729834\\
0.58	0.388986959542929\\
0.59	0.40575194658648\\
0.6	0.42335789855403\\
0.61	0.441868210458234\\
0.62	0.461352828036064\\
0.63	0.48188911640473\\
0.64	0.503562870708592\\
0.65	0.526469496592035\\
0.66	0.55071539474931\\
0.67	0.576419591931349\\
0.68	0.603715671155778\\
0.69	0.632754067174853\\
0.7	0.663704810807511\\
0.71	0.696760859548182\\
0.72	0.732141984436811\\
0.73	0.770099723980208\\
0.74	0.810923295174698\\
0.75	0.854946921833261\\
0.76	0.902558909721157\\
0.77	0.954213099985283\\
0.78	1.0104433175105\\
};
\addlegendentry{\footnotesize Proper/Succ/order1};

\addplot [color=blue,solid]
  table[row sep=crcr]{0	2.73238768281339e-05\\
0.02	0.00535291126352382\\
0.04	0.010781685450182\\
0.06	0.0165923947511849\\
0.08	0.0225448951015764\\
0.1	0.0287686523978518\\
0.12	0.035284193787173\\
0.14	0.0421105594973885\\
0.16	0.0492573576146232\\
0.18	0.0567455982927502\\
0.2	0.0645959697975353\\
0.22	0.0728426182192444\\
0.24	0.0815099695113674\\
0.26	0.0906297862541425\\
0.28	0.100237122227292\\
0.3	0.110370746378488\\
0.32	0.12107371205579\\
0.34	0.13239380880554\\
0.36	0.144384338398727\\
0.38	0.157104985973261\\
0.4	0.17062274695102\\
0.42	0.185013071491152\\
0.44	0.20036127509501\\
0.46	0.216764170369551\\
0.48	0.234332174137426\\
0.5	0.253191794131257\\
0.52	0.273487905674355\\
0.54	0.295388351451806\\
0.56	0.319088200459453\\
0.58	0.344815501678078\\
0.6	0.372838558118742\\
0.62	0.403475142982561\\
0.64	0.437104761336978\\
0.66	0.474184000490357\\
0.68	0.515267325312464\\
0.7	0.561034773686174\\
0.72	0.612329860781512\\
0.74	0.670211077744129\\
0.76	0.736025999640564\\
0.78	0.811516561134731\\
0.8	0.898975304840281\\
0.82	1.00148122599752\\
0.84	1.12326860018744\\
0.86	1.2703249105592\\
0.88	1.45140318623746\\
0.9	1.67982643296846\\
0.92	1.97693611679598\\
0.94	2.37907827057555\\
0.96	2.95393068531134\\
0.98	3.84301178455851\\
1	5.40065054459958\\
};
\addlegendentry{\footnotesize Proper/Succ/order2};

\addplot [color=black!60!green,dashed,mark=o,mark options={solid}]
  table[row sep=crcr]{0	0\\
0.05	0.0135959309990331\\
0.1	0.0286466451269864\\
0.15	0.0451372250849949\\
0.2	0.0634743507604196\\
0.25	0.0838831887445178\\
0.3	0.10641203131878\\
0.35	0.131583431090259\\
0.4	0.159585937228337\\
0.45	0.190864465456758\\
0.5	0.225862975757522\\
0.55	0.264918752521078\\
0.6	0.30869759107412\\
0.65	0.357744584176951\\
0.7	0.412717894661108\\
0.75	0.47434685111299\\
0.8	0.543427929105082\\
0.85	0.620867179599887\\
0.9	0.707582224465064\\
0.95	0.804684529229637\\
1	0.91332500785676\\
};
\addlegendentry{\footnotesize Improper/Succ/order1};

\addplot [color=black!60!green,solid,mark=*,mark options={solid}]
  table[row sep=crcr]{0	0\\
0.05	0.0136242394592417\\
0.1	0.0285894453432475\\
0.15	0.0450784161146481\\
0.2	0.0632695836668549\\
0.25	0.083431302732119\\
0.3	0.105819357992434\\
0.35	0.131188961071998\\
0.4	0.157984784530975\\
0.45	0.188501804186675\\
0.5	0.222485924120337\\
0.55	0.260345193988778\\
0.6	0.302569425067704\\
0.65	0.349693113467017\\
0.7	0.402367139440923\\
0.75	0.461332768932781\\
0.8	0.527266743630748\\
0.85	0.601185473768557\\
0.9	0.684049744942107\\
0.95	0.77716209807695\\
1	0.881668358329031\\
};
\addlegendentry{\footnotesize Improper/Succ/order2};

\addplot [color=red,dashed,mark=square,mark options={solid}]
  table[row sep=crcr]{0	0\\
0.05	0.0135912881886036\\
0.1	0.0285913979306102\\
0.15	0.0452040367571334\\
0.2	0.0635020343224239\\
0.25	0.0844106597256804\\
0.3	0.106445009101028\\
0.35	0.131567981773658\\
0.4	0.159544922993274\\
0.45	0.192541021677813\\
0.5	0.225500744653448\\
0.55	0.264333314881686\\
0.6	0.308837984703939\\
0.65	0.356991894295116\\
0.7	0.408933914927868\\
0.75	0.471352150084703\\
0.8	0.532205992615016\\
0.85	0.604212311847781\\
0.9	0.683989695042317\\
0.95	0.772566276213054\\
1	0.870840764739125\\
};
\addlegendentry{\footnotesize Impoper/Sym.ext.=3/Succ/order1};

\addplot [color=red,solid,mark=square*,mark options={solid}]
  table[row sep=crcr]{0	0\\
0.05	0.0135552676293214\\
0.1	0.0284154224909061\\
0.15	0.0446939075279709\\
0.2	0.0625842146098979\\
0.25	0.0821408659962163\\
0.3	0.103676767228622\\
0.35	0.127329386954041\\
0.4	0.153303483997962\\
0.45	0.181916084636202\\
0.5	0.213338917547283\\
0.55	0.247882165277337\\
0.6	0.285949483423499\\
0.65	0.3278551621071\\
0.7	0.374023954776067\\
0.75	0.424866863297333\\
0.8	0.480918182181493\\
0.85	0.542716622921214\\
0.9	0.610900865268147\\
0.95	0.686053182005361\\
1	0.769014372221834\\
};
\addlegendentry{\footnotesize Improper/Sym.ext.=3/Succ/order2};

\end{axis}
\end{tikzpicture}%
\label{PowerMinRateB}
}
\caption{Minimum power required to fulfil certain rate demands which is assumed to be equal for all users. Rate demands for all users are assumed to be equal. We assume that the channels remain constant over three symbols. The transmitters precode three codeword symbols jointly over an extended channel. Order 1: The base station decodes the message of the 1st user firstly, Order 2: The base station decodes the message of the 2nd user firstly.}
\label{PowerMinRate}
\end{figure*}

\begin{table}
\centering
\begin{tabular}{ |c|c|c|c||c|c||c|c|  }
 \hline
 \multicolumn{4}{|c||}{Tuples on the 3D Pareto Boundary} & \multicolumn{2}{|c||}{$min.\ P_{\Sigma}$}
 & \multicolumn{2}{c|}{Power saving ratio ($\%$)}\\
 \hline
 $R_1$ & $R_2$ & $R_3$ & $P^{'}_{\Sigma}$ & $P^{\eta =1}_{\Sigma}$ & $P^{\eta =3}_{\Sigma}$ & $1-\frac{P^{\eta =1}_{\Sigma}}{P^{'}_{\Sigma}}$ & $1-\frac{P^{\eta =3}_{\Sigma}}{P^{\eta =1}_{\Sigma}}$\\
 \hline
 $1.5219$   & $0.5073$    & $0.5073$ &   $3.00$ & $2.31$ & $1.23$ & $23\%$ & $46\%$\\
 
 $1.3051$   & $0.3263$    & $1.6314$ &   $3.00$ & $2.57$ & $1.78$ & $14\%$ & $30\%$\\
 
 $1.009$   & $1.009$    & $0.5044$ &   $3.00$ & $2.15$ & $1.19$ & $28\%$ & $44\%$\\

 $0.5105$   & $1.5316$    & $0.5105$ &   $3.00$ & $2.02$ & $1.21$ & $32\%$ & $40\%$\\
 \hline
\end{tabular}
\caption{Comparison between optimal sum power solution of problem (\ref{A1}) and problem (\ref{A6}). Solving these problems successively yields optimal power allocation for the rate tuples on the Pareto boundary. Note that, $\eta$ stands for symbol extension length.}
\label{Tab1}
\end{table}

The numerical solutions for the power minimization problem subject to rate constraints is provided in Fig.~\ref{PowerMinRate}. As expected, given users' rate constraints, improper signaling achieves the same demands with less power consumption. Using symbol extensions while assuming a time-invariant channel over the extended symbol, the power can be further decreased due to inter-symbol cooperation achieved by joint beamforming. Thereby, in a time-invariant channel, a beamforming strategy which considers improper signaling over an extended symbol consumes the least power for a given QoS constraint. Notice that under weak interference, the gap between the performance gains of the investigated schemes reduces. This can be verified by comparing Fig.~\ref{PowerMinRateA} and Fig.~\ref{PowerMinRateB}. We observe that with high interference, i.e., ${\bf H}_1$, the performance gap between different transmission/reception schemes (e.g., scheme 1: proper signaling at the transmitter and successive decoding at the receiver, scheme 2: improper signaling with symbol extensions at the transmitter and successive decoding at the receiver) is higher than the case with low interference, i.e., ${\bf H}_2$.\\
In order to design an energy efficient communication, the power minimization problem is proposed to be solved for obtaining the minimum sum power that guarantees a given rate tuple on the Pareto boundary. The performance improvement by this successive optimization (i.e., first problem (\ref{A1}) , then problem (\ref{A6}) according to the solution of problem (\ref{A1})) is presented in Table~\ref{Tab1}. As shown in this table, the proposed successive optimization results in less power consumption for achieving the rate tuples on the Pareto boundary. This procedure is mathematically formulated as
\begin{align}
\{\mathbf{r',p'}\}={\rm arg}\max_{\mathbf{r,p}}\ &R_{\Sigma}(\boldsymbol{\alpha})\quad\quad
{\rm s.t.}\quad (\ref{pow})-(\ref{psd})\label{pp1}\\
\{\mathbf{Q}^{*}_j\}={\rm arg}\min_{\mathbf{Q}_j, \forall j}\ &\sum_{j}{\rm Tr}(\mathbf{Q}_j)\quad\quad
{\rm s.t.}\quad \mathbf{r}  \geq \mathbf{r'},\ (\ref{b}),(\ref{c}),\label{pp2}
\end{align}
where $\mathbf{r'}$ and $\mathbf{p'}$ is the vector of optimal rate tuple (a rate tuple on the Pareto boundary) and the corresponding power so that $P^{'}_{\Sigma}=\mathbf{1}^{T}\mathbf{p'}$. The solution of (\ref{pp2}) for a given rate tuple on the Pareto boundary (derived from (\ref{pp1})) and given symbol extension length (say $\eta$) is $P^{\eta}_{\Sigma}=\sum_{j}{\rm Tr}(\mathbf{Q}^{*}_j)$. 
\section{Extension to Multiple MAC users}\label{NR2}
In this section we discuss the benefits of improper Gaussian signaling in a general PIMAC without any limitation on the number of MAC users. The rate and power optimization procedures can be similarly formulated as in sections III and IV. Considering successive decoding and TIN at the receivers, increasing the number of MAC users naturally results in a degradation in the achievable rates per user (i.e., MAC users and the transmitter of D2D pair). Hence, we expect to fulfil the per user rate requirements with more power consumption per user (assuming the feasibility of power allocation problem). A similar argument can be made for the rate maximization problem. Compared to a PIMAC with two MAC users, the achievable rates are degraded in case of the increment in the number of MAC users.
\subsection{Numerical Results for Multiple MAC Users}
\begin{figure*}[t]
\centering
\subfigure[The minimum power per user required to satisfy the rate demands.]{
\tikzset{every picture/.style={scale=.85}, every node/.style={scale=.91}}%
\begin{tikzpicture}

\begin{axis}[%
unbounded coords=jump,
scale only axis,
xmin=0,
xmax=0.7,
xlabel={Rate demands (bits/channel use)},
xmajorgrids,
ymin=0,
ymax=4,
ylabel={$\frac{min. \sum P}{J}$ (watts)},
ymajorgrids,
legend style={at={(axis cs:0,4)},anchor=north west, draw=black,fill=white,legend cell align=left}
]
\addplot [color=blue,solid,mark=square*,mark options={solid}]
  table[row sep=crcr]{0.02	0.0142921778518624\\
0.06	0.0456912959276183\\
0.1	0.0814756285304762\\
0.14	0.122614953243501\\
0.18	0.170304719537374\\
0.22	0.226240037623015\\
0.26	0.292696943603572\\
0.3	0.372874864307619\\
0.34	0.471439391508847\\
0.38	0.5953432426123\\
0.42	0.755749378946102\\
0.46	0.971323266178296\\
0.5	1.27614372717393\\
0.54	1.73964024510273\\
0.58	2.52848055667022\\
0.62	4.16852124650959\\
0.66	9.64275935661184\\
0.7	inf\\
};
\addlegendentry{Proper, J=3};

\addplot [color=blue,dashed,mark=square,mark options={solid}]
  table[row sep=crcr]{0.02	0.0142245238959332\\
0.06	0.0447136079130317\\
0.1	0.0781990154305921\\
0.14	0.114838146966781\\
0.18	0.155238740364998\\
0.22	0.199685316944035\\
0.26	0.248478346006202\\
0.3	0.30231257906215\\
0.34	0.361645705013963\\
0.38	0.42710897534907\\
0.42	0.498860343990928\\
0.46	0.578629546826729\\
0.5	0.666498251421142\\
0.54	0.763458008174537\\
0.58	0.870859695983939\\
0.62	0.98952223153301\\
0.66	1.1207885826996\\
0.7	1.2662874489703\\
};
\addlegendentry{Improper, J=3};

\addplot [color=red,solid,mark=*,mark options={solid}]
  table[row sep=crcr]{0.02	0.0147493807894186\\
0.06	0.0506047796908553\\
0.1	0.0981356158278613\\
0.14	0.16360105079681\\
0.18	0.258738823582424\\
0.22	0.408501394311281\\
0.26	0.677035182909326\\
0.3	1.29394823242726\\
0.34	4.15941499091273\\
0.38	inf\\
0.42	inf\\
0.46	inf\\
0.5	inf\\
0.54	inf\\
0.58	inf\\
0.62	inf\\
0.66	inf\\
0.7	inf\\
};
\addlegendentry{Proper, J=7};

\addplot [color=red,dashed,mark=o,mark options={solid}]
  table[row sep=crcr]{0.02	0.0147046258721827\\
0.06	0.0486444226363952\\
0.1	0.0899121796259561\\
0.14	0.140315918677459\\
0.18	0.201947558272874\\
0.22	0.277130709773617\\
0.26	0.368914083518002\\
0.3	0.481001586539103\\
0.34	0.617810184395888\\
0.38	0.784500645829984\\
0.42	0.987492421540759\\
0.46	1.23448567213945\\
0.5	1.53496311844421\\
0.54	1.89994656290305\\
0.58	2.34342697465016\\
0.62	2.88212131160355\\
0.66	3.53465376652432\\
0.7	4.32723188381468\\
};
\addlegendentry{Improper, J=7};

\end{axis}
\end{tikzpicture}%
\label{fig:MMU1}
}
\subfigure[The ratio of minimum sum power consumption in case of improper signaling to proper signaling]{
\tikzset{every picture/.style={scale=.85}, every node/.style={scale=.8}}%
\begin{tikzpicture}

\begin{axis}[%
unbounded coords=jump,
scale only axis,
xmin=0,
xmax=0.7,
xlabel={Rate demands (bits/channel use)},
xmajorgrids,
ymin=0,
ymax=1,
ylabel={$\frac{min. \sum P_{Improper}}{min. \sum P_{Proper}}$},
ymajorgrids,
legend style={at={(axis cs: 0, 1)},anchor=north west, draw=black,fill=white,legend cell align=left}
]
\addplot [color=blue,solid,mark=square*,mark options={solid}]
  table[row sep=crcr]{0	0.995266364816445\\
0.06	0.978602313750623\\
0.1	0.95978413227388\\
0.14	0.936575384396413\\
0.18	0.911535163480482\\
0.22	0.882625900534775\\
0.26	0.848927026524542\\
0.3	0.810761485957242\\
0.34	0.76710964660062\\
0.38	0.717416348718368\\
0.42	0.660087004883276\\
0.46	0.595712639627553\\
0.5	0.522275224356685\\
0.54	0.438859706955935\\
0.58	0.344420167157932\\
0.62	0.237379678072065\\
0.66	0.116231105770684\\
0.7	0\\
};
\addlegendentry{J=3};

\addplot [color=black!60!green,solid,mark=o,mark options={solid}]
  table[row sep=crcr]{0	0.994443579130963\\
0.06	0.974056276237031\\
0.1	0.951412998676399\\
0.14	0.921406493223696\\
0.18	0.885810800482791\\
0.22	0.844107331310572\\
0.26	0.794283755051158\\
0.3	0.734912262762377\\
0.34	0.663312174829727\\
0.38	0.578739544311007\\
0.42	0.478342437790692\\
0.46	0.359333336237594\\
0.5	0.218757186559765\\
0.54	0.0531335856834309\\
0.58	0\\
0.62	0\\
0.66	0\\
0.7	0\\
};
\addlegendentry{J=4};

\addplot [color=black,solid,mark=asterisk,mark options={solid}]
  table[row sep=crcr]{0	0.990266602302863\\
0.06	0.962488853943052\\
0.1	0.927394008892696\\
0.14	0.884556937675099\\
0.18	0.833144369887007\\
0.22	0.770321716065246\\
0.26	0.694735586482146\\
0.3	0.603836577525999\\
0.34	0.494692368163788\\
0.38	0.364390964823913\\
0.42	0.209370998101148\\
0.46	0.0251034458176308\\
0.5	0\\
0.54	0\\
0.58	0\\
0.62	0\\
0.66	0\\
0.7	0\\
};
\addlegendentry{J=5};

\addplot [color=cyan,solid,mark=square,mark options={solid}]
  table[row sep=crcr]{0	0.992940897561833\\
0.06	0.967854083058615\\
0.1	0.934370976103103\\
0.14	0.89007964904607\\
0.18	0.829990552499761\\
0.22	0.751572838178299\\
0.26	0.650931910421403\\
0.3	0.522327355441119\\
0.34	0.359812161326197\\
0.38	0.156279128894047\\
0.42	0\\
0.46	0\\
0.5	0\\
0.54	0\\
0.58	0\\
0.62	0\\
0.66	0\\
0.7	0\\
};
\addlegendentry{J=6};

\addplot [color=red,solid,mark=*,mark options={solid}]
  table[row sep=crcr]{0	0.996965640939445\\
0.06	0.961261424979302\\
0.1	0.916203346435102\\
0.14	0.857671255741075\\
0.18	0.780507368305867\\
0.22	0.678408234666737\\
0.26	0.544896473374869\\
0.3	0.371731708027309\\
0.34	0.148532951327445\\
0.38	0\\
0.42	0\\
0.46	0\\
0.5	0\\
0.54	0\\
0.58	0\\
0.62	0\\
0.66	0\\
0.7	0\\
};
\addlegendentry{J=7};

\end{axis}
\end{tikzpicture}%
\label{fig:MMU2}
}
\caption{ The performance comparison of proper and improper Gaussian signaling in the PIMAC with multiple MAC users. The rate demands per user is assumed to be equal.}
\label{fig:MMU}
\end{figure*}
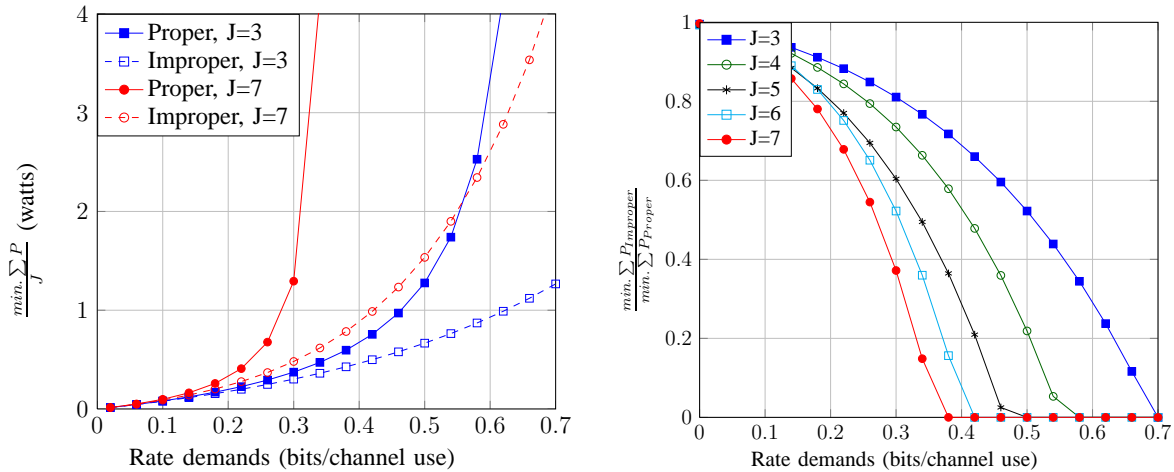

We present the numerical results for a PIMAC with 2 upto 6 MAC users (i.e., $J=3$ upto $J=7$). The channel between any communication pair (i.e., $\{h_{ij}|i\in\{1,2\},\ j\in\mathcal{J}\}$)  is given by $\mathbf{H}=\begin{bmatrix}
\mathbf{H}_1 & \mathbf{H}^{'}
\end{bmatrix} 
$, where $\mathbf{H}^{'}$ is
\begin{align}
\mathbf{H}^{'}=\begin{bmatrix}
 0.40e^{i1.3972} &  1.12e^{i0.7737}  &   0.43e^{i1.2874}  
 &  0.84e^{i0.3067}\\
 1.24e^{-i0.9872} & 1.70e^{i0.9784} &  0.83e^{-i0.2156}
  & 0.67e^{-i1.6414}
\end{bmatrix}.
\end{align}
By utilizing successive decoding with a fixed decoding order and TIN at the receivers, we compare the performance of proper and improper Gaussian signaling from the minimum power consumption perspective. As shown in Fig.~\ref{fig:MMU1}, the minimum power required per user to satisfy a particular rate demand increases with the number of MAC users. This is intuitive due to the additional interference terms in the received signal, \eqref{eq:R1}, \eqref{eq:R2}. The ratio of minimum power consumption by improper Gaussian signaling to the minimum power consumption by proper Gaussian signaling is shown in Fig.~\ref{fig:MMU2}. This ratio tends to decrease by the number of MAC users for the feasible rate demands, (e.g., when the rate constraint per user is 0.3 bits/channel~use, improper Gaussian signaling results in $20\%$ and $50\%$ reduction in sum power in PIMACs with 2-user and 6-user MAC, respectively). It is important to note that, proper Gaussian signaling does not yield a feasible power allocation solution at specific rate demands, meanwhile improper Gaussian signaling satisfies these demands. This is notable from Fig.~\ref{fig:MMU2}, where the ratio becomes zero (i.e., denominator of the ratio becomes infeasible). 

\section{Conclusion}
\label{Sec:Conc}
We investigated the achievable rate region of the MAC in the presence of interference from a point-to-point (P2P) communication system sharing the same resources, using general (improper) Gaussian transmission. This P2P system might be an underlay cognitive radio, for example. The achievable rate region is maximized with respect to the variance and pseudo-variance of the transmit signal, while treating interference as noise at the receivers. The benefit of using improper signaling is reflected by the fact that a non-zero pseudo-variance achieves a larger rate region in the MAC for a given rate of the P2P channel. Similarly, the P2P channel obtains a higher rate for a given rate of the MAC channel when using improper signaling compared to proper signaling. We also considered power minimization using different receiver structures and also using symbol extensions. In this case, improper signaling allows achieving the desired QoS while expending less power at the transmitters. Moreover, we investigated the benefits of successive rate maximization and power minimization for an energy efficient communication system design.

\appendix
We define the following real-valued vectors:
\begin{align}
\bold{c}&=[C_{x_1} \quad C_{x_2} \quad C_{x_3}]^{T},\\
\bold{a}_1&=[|h_{11}|^{2} \quad 0 \quad |h_{13}|^{2}]^{T},\\
\bold{a}_2&=[0 \quad |h_{12}|^{2} \quad |h_{13}|^{2}]^{T},\\
\bold{a}_3&=[|h_{21}|^{2} \quad |h_{22}|^{2} \quad |h_{23}|^{2}]^{T},\\
\bold{a}_4&=[|h_{11}|^{2} \quad |h_{12}|^{2} \quad |h_{13}|^{2}]^{T},\\
\bold{b}_1&=\bold{b}_2=\bold{b}_4=[0 \quad 0 \quad |h_{13}|^{2}]^{T},\\
\bold{b}_3&=[|h_{21}|^{2} \quad |h_{22}|^{2}\quad 0]^{T},
\end{align}
and the set of complex-valued vectors as
\begin{align}
\bold{\tilde{c}}&=[\tilde{C}_{x_1} \quad \tilde{C}_{x_2} \quad \tilde{C}_{x_3}]^{T},\\
\bold{\tilde{a}}_1&=[h_{11}^{2} \quad 0 \quad h_{13}^{2}]^{T},\\
\bold{\tilde{a}}_2&=[0 \quad h_{12}^{2} \quad h_{13}^{2}]^{T},\\
\bold{\tilde{a}}_3&=[h_{21}^{2} \quad h_{22}^{2} \quad h_{23}^{2}]^{T},\\
\bold{\tilde{a}}_4&=[h_{11}^{2} \quad h_{12}^{2} \quad h_{13}^{2}]^{T},\\
\bold{\tilde{b}}_1&=\bold{\tilde{b}}_2=\bold{\tilde{b}}_4=[0 \quad 0 \quad h_{13}^{2}]^{T},\\
\bold{\tilde{b}}_3&=[h_{21}^{2} \quad h_{22}^{2}\quad 0]^{T}.
\end{align}

Using these vectors, we can reformulate the optimization problem (\ref{A2}) as follows,
\begin{subequations}\label{A7}
\begin{align}
\max_{\bf{c}\rm\in\mathbb{R}^{3},\tilde{\bf{c}}\in\mathbb{C}^{3}} &\min_{q} \frac{1}{2\alpha_q}\log{\frac{({\sigma^{2}+{\bf{a}}_q^{T}
{\bf{c}}})^{2} -{\bf{\tilde{c}}}^{H} \tilde{{\bf{{A}}}}_q {\bf{\tilde{c}}} }{({\sigma^{2}+{\bf{b}}_q^{T}
{\bf{c}}})^{2} -{\bf{\tilde{c}}}^{H} \tilde{{\bf{B}}}_q {\bf{\tilde{c}}}}}\tag{\ref{A7}}\label{AppOpt1}\\
{\rm{s.t.}}\quad &{\bf{c}}{\bf{E}}_j{\bf{c}}^{T}  \leq{P_j^{2}},\quad\forall{j}\in\mathcal{J}\\
&{\bf{e}}_j^{T}{\bf{c}} \geq{0},\quad\forall{j}\in\mathcal{J}\\
&{\tilde{\bf{c}}}{\bf{E}}_j{\tilde{\bf{c}}}^{T}  \leq {\bf{c}}{\bf{E}}_j{\bf{c}}^{T},\quad\forall{j}\in\mathcal{J}
\end{align}
\end{subequations}
where, $\tilde{{\bf{{A}}}}_q={\bold{\tilde{a}}}_q
{\bold{\tilde{a}}^{T}}_q$ and $\tilde{{\bf{{B}}}}_q=\bold{\tilde{b}}_q
\bold{\tilde{b}}^{T}_q$. Note that ${\bf{e}}_j$ denotes the $j^{th}$ column of the $3\times{3}$ identity matrix and ${\bf{E}}_j$ is defined to be ${\bf{e}}_j{\bf{e}}_j^{T}$.

Optimization problem (\ref{AppOpt1}) is a non-homogeneous quadratically constraint quadratic program (QCQP).
Homogenizing the problem requires introducing another parameter \cite{Luo2010}. Thus, the homogeneous QCQP can be recast as
\begin{subequations}\label{A8}
\begin{align}
\max_{t\in\mathbb{R}, \bf{c}\rm\in\mathbb{R}^{3},\tilde{\bf{c}}\in\mathbb{C}^{3}} &\min_{q} \frac{1}{2\alpha_q}\log{\frac{({\sigma^{2}t+{\bf{a}}_q^{T}
{\bf{c}}})^{2} -{\bf{\tilde{c}}}^{H} \tilde{{\bf{{A}}}}_q {\bf{\tilde{c}}} }{({\sigma^{2}t+{\bf{b}}_q^{T}
{\bf{c}}})^{2} -{\bf{\tilde{c}}}^{H} \tilde{{\bf{B}}}_q {\bf{\tilde{c}}}}}\tag{\ref{A8}}\label{QCQP1}\\
{\rm{s.t.}}\quad &{\bf{c}}{\bf{E}}_j{\bf{c}}^{T}  \leq{P_j^{2}},\quad\forall{j}\in\mathcal{J}\\
&{\bf{e}}_j^{T}{\bf{c}}t \geq{0},\quad\forall{j}\in\mathcal{J}\\
&{\tilde{\bf{c}}}{\bf{E}}_j{\tilde{\bf{c}}}^{T}  \leq {\bf{c}}{\bf{E}}_j{\bf{c}}^{T},\quad\forall{j}\in\mathcal{J}\\
&t^{2}=1.
\end{align}
\end{subequations}
It turns out that, if the optimum values for the parameters of the homogenized optimization problem is $({\bf{c}^{*}},{\bf\tilde{c}^{*}},t^{*})$, then the optimum values of the parameters of the original optimization problem are,
${\bf{c}}_{org}^{*}={\bf{c}^{*}}/{t^{*}},\quad{\bf\tilde{c}^{*}}_{org}={\bf\tilde{c}^{*}}/{t^{*}}$ .
Now, We introduce a set of matrices as, ~\cite{Zeng2013}
\begin{align}
{\bf{C}}&=\begin{bmatrix}t\\{\bf{c}}\end{bmatrix}
\begin{bmatrix}t\\{\bf{c}}\end{bmatrix}^{T},\\
\tilde{\bf{C}}&=\tilde{\bf{c}}\tilde{\bf{c}}^{H}, \\
{\bf{W}}_q&=\begin{bmatrix}{\sigma^{2}}\\{\bf{a}}_q\end{bmatrix}
\begin{bmatrix}{\sigma^{2}}\\{\bf{a}}_q\end{bmatrix}^{T},\\
{\bf{Z}}_q&=\begin{bmatrix}{\sigma^{2}}\\{\bf{b}}_q\end{bmatrix}
\begin{bmatrix}{\sigma^{2}}\\{\bf b}_q\end{bmatrix}^{T},\\
{\bf{N}}_j&=\begin{bmatrix} 0 & 0.5{\bf{e}}_j^{T} \\ 0.5{\bf{e}}_j & 0 \end{bmatrix},\\
{\bf{M}}_j&=\begin{bmatrix} 0 & {\bf{0}} \\ {\bf{0}} & {\bf{E}}_j \end{bmatrix}.
\end{align}
By utilizing these matrices, we can reformulate the homogenized optimization problem as follows,
\begin{subequations}\label{A9}
\begin{align}
\max_{{\bf{C}}\in\mathbb{S}^{4},\tilde{\bf{C}}\in\mathbb{H}^{3}}
&\min_{q} \frac{1}{2\alpha_q}\log{\frac{{\rm{Tr}}({{\bf{W}}_q} {\bf{C}})-{\rm{Tr}}({\tilde{\bf{A}}_q} \tilde{\bf{C}})}{{{\rm{Tr}}({\bf{Z}}_q {\bf{C}})-{\rm{Tr}}({\tilde{\bf{B}}_q} \tilde{\bf{C}})}}}\tag{\ref{A9}}\label{final1}\\
{\rm{s.t.}}\quad &{\rm{Tr}}({\bf{M}}_j {\bf{C}}) \leq {P_j^{2}}, \quad \forall {j}\in\mathcal{J},\\
&{\rm{Tr}}({\bf{N}}_j {\bf{C}}) \geq 0, \quad \forall {j}\in\mathcal{J},\\
&{\rm{Tr}}({\bf{E}}_j \tilde{\bf{C}}) \leq {\rm{Tr}}({\bf{M}}_j {\bf{C}}), \quad \forall {j}\in\mathcal{J},\\
&{\bf{C}}_{11}=1,\label{C11} \\
&{\bf{C}}\succeq{0},\quad
\tilde{\bf{C}}\succeq{0},\\
&\rm{rank}({\bf{C}})=1, \quad \rm{rank}({\tilde{\bf{C}}})=1\label{rarank1}.
\end{align}
\end{subequations}
Based on the fact that the cone of rank-1 semidefinite matrices is not convex \cite{Boyd2004}, we relax rank-1 constraints in order to fall into the convex problem. It is important to note that the relaxed problem might end up in a solution that is far from the solution of the original problem due to the relaxation.

The constraint (\ref{C11}), i.e., ${\bf{C}}_{11}=1$,  justifies $t^{2}=1$ in the homogenized quadratic problem of (\ref{QCQP1}), where  ${\bf{C}}_{11}$ is the element in the $1^{st}$ row and the $1^{st}$ column of ${\bf{C}}$.

We can distinguish that by reformulating the homogenized problem to the SDP, the quadratic terms are converted to linear terms. Bare in mind that the optimal $\bf{c}$ and $\bf\tilde{c}$ are $3\times{1}$ vectors, but in the equivalent SDP, the search space for $\bf{C}$ and $\bf\tilde{C}$ expands to the semidefinite cone of $4\times{4}$ and $3\times{3}$ matrices, respectively.
The conditions,
\begin{align}
{\rm{Tr}}({\bf{W}}_q {\bf{C}})-{\rm{Tr}}(\tilde{\bf{A}}_q \tilde{\bf{C}})&\geq{\sigma^{4}}, \quad\forall{q}\in\mathcal{J}\cup{\{4\}},\\
{\rm{Tr}}({\bf{Z}}_q {\bf{C}})-{\rm{Tr}}(\tilde{\bf{B}}_q \tilde{\bf{C}})&\geq{\sigma^{4}}, \quad\forall{q}\in\mathcal{J}\cup{\{4\}}.
\end{align}
are always fulfilled in (\ref{A9}),~\cite{Zeng2013}. The strict positivity of these conditions in the relaxed problem (problem (\ref{A9}) without constraint (\ref{rarank1})), converts the problem into a quasi-convex problem which can be solved by the bisection method~\cite{Zeng2013}. Thus, we include these inequality constraints in the relaxed problem.
Hence, the relaxed semidefinite program can be recast as the following feasibility problem,
\begin{subequations}\label{A10}
\begin{align}
\rm{find}\quad &{\bf{C}}\in\mathbb{S}^{4}, \tilde{\bf{C}}\in\mathbb{H}^{3}\tag{\ref{A10}}\\
{\rm{s.t.}}\quad &{\rm{Tr}}({\bf{M}}_j {\bf{C}}) \leq {P_j^{2}}, \quad \forall {j}\in\mathcal{J},\\
&{\rm{Tr}}({\bf{N}}_j {\bf{C}}) \geq 0, \quad \forall {j}\in\mathcal{J},\\
&{\rm{Tr}}({\bf{E}}_j \tilde{\bf{C}}) \leq {\rm{Tr}}({\bf{M}}_j {\bf{C}}), \quad \forall {j}\in\mathcal{J},\\
&{\bf{C}}_{11}=1, \\
&{\bf{C}}\succeq{\bf{0}},\quad
\tilde{\bf{C}}\succeq{\bf{0}},\\
&{\rm{Tr}}({\bf{W}}_q {\bf{C}})-{\rm{Tr}}(\tilde{\bf{A}}_q \tilde{\bf{C}})\geq{\sigma^{4}}, \quad\forall{q}\in\mathcal{J}\cup{\{4\}},\\
&{\rm{Tr}}({\bf{Z}}_q {\bf{C}})-{\rm{Tr}}(\tilde{\bf{B}}_q \tilde{\bf{C}})\geq{\sigma^{4}}, \quad\forall{q}\in\mathcal{J}\cup{\{4\}},\\
\label{Merg11}
&{\rm{Tr}}({\bf{W}}_q {\bf{C}})-{\rm{Tr}}(\tilde{\bf{A}}_q \tilde{\bf{C}})\geq{e^{2\alpha_q R}}({\rm{Tr}}({\bf{Z}}_q {\bf{C}})-{\rm{Tr}}(\tilde{\bf{B}}_q \tilde{\bf{C}}))\quad \forall{q}\in\mathcal{J}\cup{\{4\}},
\end{align}
\end{subequations}
where the objectives are incorporated into the constraints by introducing an auxiliary optimization parameter $R$ (constraint (\ref{Merg11})).
This feasibility problem can be solved by bisection over $R$ in order to achieve the maximum sum rate in a given direction $\boldsymbol{\alpha}$ on the rate region. Solving the feasibility problem in different traverse directions will yield the achievable rate region. Note that, $\mathbb{S}^{4}$ and $\mathbb{H}^{3}$ are the set of $4\times 4$ positive semidefinite symmetric matrices and $3\times 3$ positive semidefinite Hermitian matrices, respectively.
Denoting the optimal solution for the relaxed problem by $(\bf{C^{*}},\bf\tilde{C}^{*})$, we use the well-known Gaussian randomization procedure~\cite{Luo2010},~\cite{Quan2006},~\cite{Sidi2006},\cite{Quan2007} in order to project the solutions onto the rank-1 positive semidefinite set. Note that, the randomization procedure ends up in an approximate solution depending on the number of randomizations.

\bibliographystyle{IEEEtran} 
\bibliography{reference}
\end{document}